\documentclass[nofootinbib,final,aps,secnumarabic,twocolumn]{revtex4}%
\usepackage{amsfonts}
\usepackage{amsmath}
\usepackage{amssymb}
\usepackage{graphicx}
\usepackage[usenames]{color}%
\providecommand{\U}[1]{\protect\rule{.1in}{.1in}}

\allowdisplaybreaks
\begin{document}
\title[Finite temperature effective field theory]{Finite temperature effective field theory and two-band superfluidity in Fermi gases}
\author{Serghei N. Klimin}
\altaffiliation{Also at Department of Theoretical Physics, State University of Moldova}

\email{sergei.klimin@uantwerpen.be}
\affiliation{TQC, Universiteit Antwerpen, Universiteitsplein 1, B-2610 Antwerpen, Belgium}
\author{Jacques Tempere}
\altaffiliation{Also at Lyman Laboratory of Physics, Harvard University}

\affiliation{TQC, Universiteit Antwerpen, Universiteitsplein 1, B-2610 Antwerpen, Belgium}
\author{Giovanni Lombardi}
\affiliation{TQC, Universiteit Antwerpen, Universiteitsplein 1, B-2610 Antwerpen, Belgium}
\author{Jozef T. Devreese}
\altaffiliation{Also at Technische Universiteit Eindhoven}

\affiliation{TQC, Universiteit Antwerpen, Universiteitsplein 1, B-2610 Antwerpen, Belgium}

\begin{abstract}
We develop a description of fermionic superfluids in terms of an effective
field theory for the pairing order parameter. Our effective field theory
improves on the existing Ginzburg - Landau theory for superfluid Fermi gases
in that it is not restricted to temperatures close to the critical
temperature. This is achieved by taking into account long-range fluctuations
to all orders. The results of the present effective field theory compare well
with the results obtained in the framework of the Bogoliubov - de Gennes
method. The advantage of an effective field theory over Bogoliubov - de Gennes
calculations is that much less computation time is required. In the second
part of the paper, we extend the effective field theory to the case of a
two-band superfluid. The present theory allows us to reveal the presence of
two healing lengths in the two-band superfluids, to analyze the
finite-temperature vortex structure in the BEC-BCS crossover, and to obtain
the ground state parameters and spectra of collective excitations. For the
Leggett mode our treatment provides an interpretation of the observation of
this mode in two-band superconductors.

\end{abstract}
\date{\today}
\maketitle

\section{Introduction}

Multi-bandgap superconductivity, predicted by Suhl, Matthias, and Walker
\cite{SuhlPRL3}, was first revealed in MgB$_{2}$ \cite{NagamatsuNAT,LiuPRL87},
and more recently in the iron pnictide class of superconductors
\cite{KamiharaJACS128}. The multiple bandgaps arise from differences in
character between the Fermi surface sheets on which Cooper pairing takes place
\cite{LiuPRL87}. In the two-bandgap superconductor MgB$_{2}$, the two Cooper
pairing channels moreover appear to be in different regimes: taken
individually they would lead to type I and type II superconductivity
respectively. Therefore, this material was dubbed a \textquotedblleft type
1.5\textquotedblright\ superconductor \cite{MoshchalkovPRL102}. The competing
length scales associated with the Cooper pairing channels lead to the
formation of vortex clusters and stripes
\cite{MoshchalkovPRL102,GutierrezPRB85}. The experimental discovery of vortex
clustering in MgB$_{2}$ has lead to a flurry of activity to develop a
two-bandgap Ginzburg - Landau (GL) formalism suitable to describe these patterns.

The increasing interest in two-band superfluid fermionic system is not
restricted to superconductors \cite{BabaevNAT431}. Recently, the superfluidity
of multiband ultracold atomic Fermi gases has attracted theoretical attention
\cite{Iskin,Iskin2,Iskin3}, anticipating interesting experiments in this
field. Quantum gases offer the singular advantage that the adaptability of
various experimental parameters (intraband and interband interaction strength,
numbers of atoms, trapping geometry,...) allows to study these systems in
regimes inaccessible in solids. A GL theory has been developed for these
systems at the microscopic level \cite{Iskin3,SadeMeloPRL71,HuangPRA79}, as
distinct from the case of superconductivity where many parameters remain
phenomenological. Here, we focus on two-bandgap superfluidity in atomic Fermi
gases throughout the crossover from the weak-coupling BCS regime to the
Bose-Einstein condensate (BEC) regime, where pairing of molecules in real
space occurs.

In the straightforward two-component GL expansion (TCGL) two single-component
GL equations are coupled through a Josephson term (see, e. g. Refs.
\cite{ZhitomirskyPRB69,Gurevich2003,BabaevPRB72}), and lead to an intervortex
interaction that can account for vortex clustering \cite{BabaevPRL105}.
However, the validity of this simple extension has been the subject of intense
debate
\cite{KoganPRB83,KoganPRB86,BabaevPRB86,ShanenkoPRL106,Shanenko2,Babaev2011,Babaev2012-2}%
. Kogan and Schmalian \cite{KoganPRB83,KoganPRB86} indicate that the two order
parameters in a two-band superconductor should have the same length scale of
spatial variation in the vicinity of the critical temperature $T_{c}$, when
$T\rightarrow T_{c}$. Since the standard GL formalism is developed for $T$
near $T_{c}$, these authors conclude that the GL approach fails to adequately
describe the existence of two different length scales in a two-band
superconductor. On the other hand, Babaev and Silaev \cite{BabaevPRB86} argue
that the TCGL expansion is justified and properly describes two-band systems
with different coherence lengths. Both sides, however, recognize that the
temperature range of validity for the TCGL approach is restricted from below
by the condition that the order parameter amplitude is small
\cite{Babaev2012-2}. Therefore, finding an effective TCGL-like formalism valid
well below $T_{c}$ remains an open question. In Refs.
\cite{ShanenkoPRL106,Shanenko2,Sh3,Orlova}, an extended two-component GL
formalism is found by performing an expansion of the free energy and the gap
equation in powers of $\tau=1-T/T_{c}\ $to order $\tau^{3/2}$ rather than
$\tau^{1/2}$ as is common for the standard GL formalism. This approach
confirms the existence of two distinct length scales \cite{Sh4}. However in
practice a complete summation of the series over $\tau$ is not feasible.

It was shown \cite{Babaev2012-2} that a TCGL model with phenomenologically
determined coefficients yields an accurate description of vortices and of the
magnetic response of a two-band superconductor in a wide range of
temperatures. Models where the GL parameters are calculated from a microscopic
theory are available in the limit of weak-coupling BCS superconductors (e. g.
Refs. \cite{Tewordt,Werthammer}), where the assumption of slowly varying
fields was a key ingredient. Here, we invoke the same assumption to develop a
theory that avoids any additional approximation (for example, small $\tau$,
small pair field, or weak coupling) and that retrieves in limiting cases the
results of known effective field theories. Our finite-temperature effective
field theory retrieves the zero-temperature effective field theory
\cite{Marini1998,Schakel} in the limit $T\rightarrow0$ throughout the BCS-BEC
crossover. Also in the other limit, $T\rightarrow T_{c}$, the obtained EFT
analytically reproduces the results obtained by the microscopic path-integral
treatment for the homogeneous superfluid in the entire BCS-BEC crossover
\cite{SadeMeloPRL71,HuangPRA79}. The effective field theory that we obtain in
this way has been applied successfully to dark solitons in ultracold Fermi
gases \cite{DSol}, where it shows a good agreement with Bogoliubov - de Gennes
theory. The present work for the first time systematically describes the
derivation of the finite temperature EFT formalism, which is only briefly
represented in Ref. \cite{DSol}, and applies the theory to describe vortex
structure in the BCS-BEC crossover.

Next, we extend the effective field theory to interacting mixtures of
superfluid Fermi gases. When two pairing channels are available, these systems
represent the quantum gas analog of the two-band superconductors discussed
above. Specifying the species of trapped atoms, their hyperfine states, and
the number of trapped atoms, fixes unambiguously the microscopic Hamiltonian
in terms of scattering lengths, chemical potentials, and masses. Starting from
the microscopic action functional for two-band atomic Fermi gases with
$s$-wave pairing, we obtain unique expressions for the parameters of the
effective field theory for the two band superfluid, including expressions for
the Josephson coupling between the two order parameters as a function of the
scattering lengths. The resulting effective field theory reveals the presence
of two healing length scales in the two-band superfluids, in close analogy to
the so-called hidden criticality discussed for two-band superconductors. In
order for the theory to be capable of describing the experimentally relevant
collective excitations of superfluid Fermi gases, a derivative expansion
keeping only the first order derivatives of the pair field over time
(performed, e.~g. in Refs. \cite{SadeMeloPRL71,HuangPRA79}) is not sufficient:
second-order time derivatives are required to determine collective excitation
spectra of Fermi superfluids. Including these second-order derivatives, we
obtain the collective modes including the Leggett mode, and compare the
results obtained in the framework of superfluid two-band systems to
experimental results obtained for the Leggett mode in two-band superconductors.

The paper is divided in two parts. In the first part, Sect.~\ref{EFT}, we
derive the effective field action and the field equations for a
single-component Fermi superfluid (subsection \ref{subsec2a}). In this part we
also compare the results for the thermodynamics of the uniform system to the
results of the microscopic description to show the validity of the field
theory for a large temperature range in subsection \ref{subsec2b}. Also the
structure of a vortex in the BCS-BEC crossover (subsect. \ref{subsec2c}), as
well as the collective excitation spectrum (subsect. \ref{subsec2d}) are
calculated and compared to existing treatments such as the Bogoliubov - de
Gennes treatment. In the second part (Sect. \ref{TwoBand}), we extend the
results to a two-band system (subsect. \ref{subsec3a}), and consider the
behavior of the parameters and thermodynamic quantities of two-band superfluid
Fermi gases at zero temperature and at finite temperatures (subsect.
\ref{subsec3b}). The spectra of collective excitations are again calculated
(subsect. \ref{subsec3c}), revealing for the two-band case also the Leggett
mode, i.e. the out-of-phase oscillation mode between the two bands. The
discussion is summarized in Conclusions, Sect.~\ref{Conclusions}.

\section{Effective field theory for superfluid Fermi gases \label{EFT}}

\subsection{Derivation of the field equations \label{subsec2a}}

\subsubsection{Functional integral formalism}

An effective field theory for the superfluid order parameter constitutes a
powerful tool to study non-uniform phenomena in fermionic superfluids, such as
vortices, solitons, and the effects of strong confinement. Examples are the
Gross-Pitaevskii equation for the temperature-zero Bose gas and the Ginzburg -
Landau theory for superconductors near the critical temperature. These
approaches are complementary to microscopic descriptions such as the
Bogoliubov - de Gennes approach. The latter works well for small number of
particles, whereas a description in terms of an effective field theory meets
no difficulties for large numbers of particles$,$ including the thermodynamic
limit. The other advantage of an effective-field based description is that
this usually requires much less computation time and memory than the
Bogoliubov - de Gennes calculation. Up to now, Ginzburg - Landau (GL) type
effective field theories have been developed for superfluid Fermi gases at
$T\approx T_{c}$ \cite{SadeMeloPRL71,HuangPRA79} or at $T=0$
\cite{Marini1998,Schakel}. Both assume a slow variation of a pair field in
space and time, and account for amplitude as well as phase field fluctuations.
For the two-dimensional Fermi superfluid, a finite-temperature effective field
theory has been formulated taking into account phase fluctuations in 2D
\cite{Babaev,Botelho2006}. An effective field theory for cold Fermi gases in
3D has been derived within the mean-field approximation \cite{Hsiang}. The
goal of the first part of the present paper is to develop an effective field
theory that is valid in the whole temperature range up to $T_{c}$ and accounts
for both amplitude and phase of the pair field \emph{without assuming
fluctuations small}. This extension in performed within the functional
integral formalism used in Ref. \cite{SadeMeloPRL71} and in subsequent works.
No additional hypotheses or modelling are introduced.

We consider a fermionic system of particles with two spin states each
($\sigma=\uparrow,\downarrow$). In the functional integral formalism, the
partition function of the fermionic system is determined by the path integral
over the fermion fields (the Grassmann variables):%
\begin{equation}
\mathcal{Z}\propto\int\mathcal{D}\left[  \bar{\psi},\psi\right]  e^{-S}.
\end{equation}
The system is described by the action functional $S$ of the fermionic fields
$\psi_{\sigma}$, which is given by%
\begin{equation}
S=S_{0}+\int_{0}^{\beta}d\tau\int d\mathbf{r}~U\left(  \mathbf{r},\tau\right)
, \label{S1}%
\end{equation}
where $\beta=1/\left(  k_{B}T\right)  $, $T$ is the temperature, $k_{B}$ is
the Boltzmann constant, and $S_{0}$ is the free-fermion action,%
\begin{equation}
S_{0}=\int_{0}^{\beta}d\tau\int d\mathbf{r}\sum_{\sigma=\uparrow,\downarrow
}\bar{\psi}_{\sigma}\left(  \frac{\partial}{\partial\tau}+H_{\sigma}\right)
\psi_{\sigma}. \label{S01}%
\end{equation}
The one-particle Hamiltonian $H_{\sigma}=-\nabla_{\mathbf{r}}^{2}%
/(2m)-\mu_{\sigma}$ allows for population imbalance through the chemical
potentials $\mu_{\sigma}$. The interaction Hamiltonian $U\left(
\mathbf{r},\tau\right)  $ describes the contact interactions between fermions:%
\begin{equation}
U=g\bar{\psi}_{\uparrow}\bar{\psi}_{\downarrow}\psi_{\downarrow}\psi
_{\uparrow} \label{U1}%
\end{equation}
The interaction energy with the coupling constant $g$ is determined by the
$s$-wave scattering between two fermions with antiparallel spins: this is the
Cooper pairing channel. We use the following set of units: $\hbar=1$, $m=1/2$,
and the Fermi energy for a free-particle Fermi gas $E_{F}\equiv\hbar^{2}%
k_{F}^{2}/\left(  2m\right)  =1$, where $k_{F}\equiv\left(  3\pi^{2}n\right)
^{1/3}$ is the Fermi wave vector and $n$ is the fermion particle density. The
antisymmetry requirement for fermionic wave functions prohibits $s$-wave
scattering between fermions with parallel spin.

The Hubbard-Stratonovich (HS) transformation is based on introducing bosonic
fields $\bar{\Psi},\Psi$ such that the partition function is represented
through the path integral over the Fermi and Bose fields,%
\begin{equation}
\mathcal{Z}\propto\int\mathcal{D}\left[  \bar{\psi},\psi\right]
\int\mathcal{D}\left[  \bar{\Psi},\Psi\right]  e^{-S_{HS}}. \label{Z1a}%
\end{equation}
The HS action which exactly decouples the four-field interaction terms in the
initial Hamiltonian, is the same as in Ref. \cite{SadeMeloPRL71},%
\begin{equation}
S_{HS}=S_{0}+S_{B}+\int_{0}^{\beta}d\tau\int d\mathbf{r}\left(  \bar{\Psi}%
\psi_{\uparrow}\psi_{\downarrow}+\Psi\bar{\psi}_{\downarrow}\bar{\psi
}_{\uparrow}\right)  , \label{SHS1a}%
\end{equation}
with the free-boson action%
\begin{equation}
S_{B}=-\int_{0}^{\beta}d\tau\int d\mathbf{r~}\frac{1}{g}\bar{\Psi}\Psi.
\label{SB1}%
\end{equation}

In order to address the whole range of the BCS-BEC crossover, the coupling
constant $g$ is renormalized through the $s$-wave scattering length $a_{s}$
exactly as in Ref. \cite{SadeMeloPRL71} for the one-band system:
\begin{equation}
\frac{1}{g}=m\left(  \frac{1}{4\pi a_{s}}-\int_{k<K}\frac{d\mathbf{k}}{\left(
2\pi\right)  ^{3}}\frac{1}{k^{2}}\right)  , \label{g}%
\end{equation}
with the ultraviolet cutoff $K\rightarrow\infty$. The integration over the
fermion fields leads to the partition function,%
\begin{equation}
\mathcal{Z}\propto\int\mathcal{D}\left[  \bar{\Psi},\Psi\right]  e^{-S_{eff}},
\label{Z}%
\end{equation}
with the effective bosonic action $S_{eff}$ depending on the pair field only:%
\begin{equation}
S_{eff}=S_{B}-\operatorname{Tr}\ln\left[  -\mathbb{G}^{-1}\right]  .
\label{Seff1a}%
\end{equation}
Here $\mathbb{G}^{-1}\left(  \mathbf{r},\tau\right)  =\mathbb{G}_{0}%
^{-1}\left(  \mathbf{r},\tau\right)  -\mathbb{F}\left(  \mathbf{r}%
,\tau\right)  $ is the inverse Nambu tensor, written as a sum of the
free-fermion inverse Nambu tensor $\mathbb{G}_{0}^{-1}$ and the matrix
$\mathbb{F}$ proportional to the pair field $\Psi$:%
\begin{align}
\mathbb{G}_{0}^{-1}\left(  \mathbf{r},\tau\right)   &  =\left(
\begin{array}
[c]{cc}%
-\frac{\partial}{\partial\tau}-\hat{H}_{\uparrow} & 0\\
0 & -\frac{\partial}{\partial\tau}+\hat{H}_{\downarrow}%
\end{array}
\right)  ,\label{G0a}\\
\mathbb{F}\left(  \mathbf{r},\tau\right)   &  =\left(
\begin{array}
[c]{cc}%
0 & -\Psi\left(  \mathbf{r},\tau\right) \\
-\bar{\Psi}\left(  \mathbf{r},\tau\right)  & 0
\end{array}
\right)  . \label{Fa}%
\end{align}
The effective action (\ref{Seff1a}) is expanded as a series in powers of the
pair field:%
\begin{equation}
S_{eff}=S_{B}-\operatorname{Tr}\ln\left[  -\mathbb{G}_{0}^{-1}\right]
+\sum_{p=1}^{\infty}\frac{1}{p}\operatorname{Tr}\left[  \left(  \mathbb{G}%
_{0}\mathbb{F}\right)  ^{p}\right]  . \label{Sp}%
\end{equation}
As the integration over the bosonic fields cannot be performed analytically
for the effective action (\ref{Sp}), approximations are necessary.

\subsubsection{Gradient expansion}

The crudest approximation would be to assume the pair field to be constant in
space and time, $\Psi\left(  \mathbf{r},\tau\right)  =\left\vert
\Psi\right\vert $, so that $\mathbb{F}\left(  \mathbf{r},\tau\right)
=\mathbb{F(}\mathbf{r}_{0},\tau_{0})=\mathbb{F}_{0}~$is independent of space
and time. This is the saddle-point approximation, and it corresponds to
replacing all factors $\mathbb{F}$ in $\left(  \mathbb{G}_{0}\mathbb{F}%
\right)  ^{p}$ by the constant $\mathbb{F}_{0}$. Then, the sum over all orders
of $p$ in expression (\ref{Sp}) can be performed analytically. One readily
obtains the saddle point action, and the corresponding saddle point free
energy
\begin{align}
\Omega_{s}\left(  w\right)   &  =-\int\frac{d\mathbf{k}}{\left(  2\pi\right)
^{3}}\left(  \frac{1}{\beta}\ln\left(  2\cosh\beta E_{\mathbf{k}}+2\cosh
\beta\zeta\right)  \right. \nonumber\\
&  \left.  -\xi_{\mathbf{k}}-\frac{w}{2k^{2}}\right)  -\frac{w}{8\pi a_{s}},
\label{Ws}%
\end{align}
with $w=\left\vert \Psi\right\vert ^{2}$. Here, $E_{\mathbf{k}}=\sqrt
{\xi_{\mathbf{k}}^{2}+w}$ is the Bogoliubov excitation energy, and
$\xi_{\mathbf{k}}=k^{2}-\mu$ is the free-fermion energy. The chemical
potentials for the imbalanced fermions are expressed as $\mu_{\uparrow}%
=\mu+\zeta$ and $\mu_{\downarrow}=\mu-\zeta$.

To improve on the saddle-point approximation, Gaussian pair fluctuations can
be taken into account. Then, one writes $\Psi\left(  \mathbf{r},\tau\right)
=\left\vert \Psi\right\vert +\delta\Psi$ with corresponding $\mathbb{F}\left(
\mathbf{r},\tau\right)  =\mathbb{F}_{0}+\delta\mathbb{F}(\mathbf{r},\tau)$,
and expands the action functional up to second order in the small parameter
$\delta\Psi$. This is equivalent to truncating the sum over $p$ in expression
(\ref{Sp}) at $p=2$. This restricted sum leads to gaussian path integrals
which can be performed analytically
\cite{SadeMeloPRL71,HuangPRA79,Marini1998,Schakel}.

In the present work, we go beyond this limitation, and again take the sum over
all powers of $p$. To do this, we assume that the pair field varies slowly in
space and time, so we can expand the matrix $\mathbb{F}$ around its background
value
\begin{equation}
\mathbb{F}\left(  \mathbf{r},\tau\right)  =\mathbb{F}_{0}+\left.
\partial_{\tau}\mathbb{F}\right\vert _{0}(\tau-\tau_{0})+\left.
\mathbf{\nabla}\mathbb{F}\right\vert _{0}\cdot\left(  \mathbf{r}%
-\mathbf{r}_{0}\right)  +\mathbf{...} \label{Fexp}%
\end{equation}
taking also second derivatives (not written down here) into account.
Subsequently, we replace all but (at most) two factors $\mathbb{F}$ in
$\left(  \mathbb{G}_{0}\mathbb{F}\right)  ^{p}$ by $\mathbb{F}_{0}$. The
remaining factors $\mathbb{F}$ in $\left(  \mathbb{G}_{0}\mathbb{F}\right)
^{p}$ are then expanded according to (\ref{Fexp}). Since the coefficients
$\left.  \mathbf{\nabla}\mathbb{F}\right\vert _{0}$, $\left.  \partial_{\tau
}\mathbb{F}\right\vert _{0}$,... are constant, we find that the trace of
$\left(  \mathbb{G}_{0}\mathbb{F}\right)  ^{p}$ can be taken and summed over
all $p$ analytically. Thus, after the expansion of the action (\ref{Sp}) in
gradients, we perform the \emph{complete} summation over $p$
\emph{analytically exactly}, without assuming $\mathbb{F}_{0}$ small.
Correspondingly, the range of applicability of this derivative expansion is
the same as for the Ginzburg - Landau approach as far as the spatial and
temporal variations are concerned, but without assuming the \textquotedblleft
background\textquotedblright\ $\Psi$ small. A similar scheme was developed in
Refs. \cite{Marini1998,Schakel} at $T=0$ and in the unitarity regime. Here, we
perform the complete summation of the series in powers of $\bar{\Psi},\Psi$ at
finite temperatures and at arbitrary coupling strengths.

\subsubsection{Effective action functional}

As a result, the effective bosonic action $S_{eff}$ is approximated by the
following effective field action $S_{EFT}$:%
\begin{align}
&  S_{EFT}=\int_{0}^{\beta}d\tau\int d\mathbf{r}\left\{  \left[  \Omega
_{s}\left(  w\right)  +\frac{\mathcal{D}\left(  w\right)  }{2}\left(
\bar{\Psi}\frac{\partial\Psi}{\partial\tau}\right.  \right.  \right.
\nonumber\\
&  \left.  -\frac{\partial\bar{\Psi}}{\partial\tau}\Psi\right)
+\mathcal{\tilde{Q}}\left(  w\right)  \frac{\partial\bar{\Psi}}{\partial\tau
}\frac{\partial\Psi}{\partial\tau}-\frac{\mathcal{R}\left(  w\right)  }%
{2w}\left(  \frac{\partial w}{\partial\tau}\right)  ^{2}\nonumber\\
&  \left.  \left.  +\frac{\mathcal{\tilde{C}}\left(  w\right)  }{2m}\left(
\nabla_{\mathbf{r}}\bar{\Psi}\cdot\nabla_{\mathbf{r}}\Psi\right)
-\frac{\mathcal{E}\left(  w\right)  }{2mw}\left(  \nabla_{\mathbf{r}}w\right)
^{2}\right]  \right\}  .\label{FGL2}%
\end{align}
The (local) saddle-point thermodynamic potential is still determined by the
modulus squared of the position-dependent order parameter, $w=\left\vert
\Psi\right\vert ^{2}$ as in expression (\ref{Ws}). The other coefficients are
given by:%
\begin{align}
\mathcal{\tilde{C}} &  =\int\frac{d\mathbf{k}}{\left(  2\pi\right)  ^{3}}%
\frac{k^{2}}{3m}f_{2}\left(  \beta,E_{\mathbf{k}},\zeta\right)  ,\label{c}\\
\mathcal{D} &  =\int\frac{d\mathbf{k}}{\left(  2\pi\right)  ^{3}}\frac
{\xi_{\mathbf{k}}}{w}\left[  f_{1}\left(  \beta,\xi_{\mathbf{k}},\zeta\right)
-f_{1}\left(  \beta,E_{\mathbf{k}},\zeta\right)  \right]  ,\label{d}\\
\mathcal{E} &  =2w\int\frac{d\mathbf{k}}{\left(  2\pi\right)  ^{3}}\frac
{k^{2}}{3m}\xi_{\mathbf{k}}^{2}~f_{4}\left(  \beta,E_{\mathbf{k}}%
,\zeta\right)  ,\label{ee}\\
\mathcal{\tilde{Q}} &  =\frac{1}{2w}\int\frac{d\mathbf{k}}{\left(
2\pi\right)  ^{3}}\left[  f_{1}\left(  \beta,E_{\mathbf{k}},\zeta\right)
\right.  \nonumber\\
&  \left.  -\left(  E_{\mathbf{k}}^{2}+\xi_{\mathbf{k}}^{2}\right)
f_{2}\left(  \beta,E_{\mathbf{k}},\zeta\right)  \right]  ,\label{qq}\\
\mathcal{R} &  =\int\frac{d\mathbf{k}}{\left(  2\pi\right)  ^{3}}\left[
\frac{f_{1}\left(  \beta,E_{\mathbf{k}},\zeta\right)  +\left(  E_{\mathbf{k}%
}^{2}-3\xi_{\mathbf{k}}^{2}\right)  f_{2}\left(  \beta,E_{\mathbf{k}}%
,\zeta\right)  }{3w}\right.  \nonumber\\
&  \left.  +\frac{4\left(  \xi_{\mathbf{k}}^{2}-2E_{\mathbf{k}}^{2}\right)
}{3}f_{3}\left(  \beta,E_{\mathbf{k}},\zeta\right)  +2E_{\mathbf{k}}^{2}%
wf_{4}\left(  \beta,E_{\mathbf{k}},\zeta\right)  \right]  .\label{rr}%
\end{align}
The functions $f_{p}\left(  \beta,\varepsilon,\zeta\right)  $ are determined
explicitly using the recurrence relations:%
\begin{align}
f_{1}\left(  \beta,\varepsilon,\zeta\right)   &  =\frac{1}{2\varepsilon}%
\frac{\sinh(\beta\varepsilon)}{\cosh(\beta\varepsilon)+\cosh(\beta\zeta
)},\label{msum}\\
f_{p+1}\left(  \beta,\varepsilon,\zeta\right)   &  =-\frac{1}{2p\varepsilon
}\frac{\partial f_{p}\left(  \beta,\varepsilon,\zeta\right)  }{\partial
\varepsilon}.
\end{align}
The corresponding equations for the order parameter follow from the stationary
action principle with the functional (\ref{FGL2}). In order to study the
evolution of the order parameter in real time, we replace $\tau$ by $it$ as in
Ref. \cite{SadeMeloPRL71}. This results in a set of two coupled equations. The
first equation reads:%
\begin{align}
&  i\frac{\partial\left(  w\mathcal{D}\right)  }{\partial w}\frac{\partial
\Psi}{\partial t}=\mathcal{A}\left(  w\right)  \Psi+\mathcal{Q}\frac
{\partial^{2}\Psi}{\partial t^{2}}-\frac{\mathcal{R}\Psi^{2}}{w}\frac
{\partial^{2}\bar{\Psi}}{\partial t^{2}}\nonumber\\
&  -\frac{1}{w}\frac{\partial\left(  w\mathcal{R}\right)  }{\partial w}%
\Psi\frac{\partial\bar{\Psi}}{\partial t}\frac{\partial\Psi}{\partial t}\\
&  +\left(  \frac{\partial\mathcal{Q}}{\partial w}+\frac{1}{2w}\frac
{\partial\left(  w\mathcal{R}\right)  }{\partial w}\right)  \bar{\Psi}\left(
\frac{\partial\Psi}{\partial t}\right)  ^{2}\nonumber\\
&  -\frac{1}{2}\frac{\partial\left(  \frac{\mathcal{R}}{w}\right)  }{\partial
w}\Psi^{3}\left(  \frac{\partial\bar{\Psi}}{\partial t}\right)  ^{2}\\
&  -\mathcal{C}\nabla_{\mathbf{r}}^{2}\Psi+\frac{2\mathcal{E}\Psi^{2}}%
{w}\nabla_{\mathbf{r}}^{2}\bar{\Psi}+\frac{2}{w}\frac{\partial\left(
w\mathcal{E}\right)  }{\partial w}\Psi\left(  \nabla_{\mathbf{r}}\bar{\Psi
}\cdot\nabla_{\mathbf{r}}\Psi\right)  \nonumber\\
&  -\left(  \frac{\partial\mathcal{C}}{\partial w}+\frac{1}{w}\frac
{\partial\left(  w\mathcal{E}\right)  }{\partial w}\right)  \bar{\Psi}\left(
\nabla_{\mathbf{r}}\Psi\right)  ^{2}+\frac{\partial\left(  \frac{\mathcal{E}%
}{w}\right)  }{\partial w}\Psi^{3}\left(  \nabla_{\mathbf{r}}\bar{\Psi
}\right)  ^{2},\label{eqmot}%
\end{align}
and the other equation is conjugate to (\ref{eqmot}). Here, the coefficients
$\mathcal{C}$ and $\mathcal{Q}$ are related to, respectively, $\mathcal{\tilde
{C}}$ and $\mathcal{\tilde{Q}}$ by:%
\begin{equation}
\mathcal{C}=\mathcal{\tilde{C}}-2\mathcal{E},\quad\mathcal{Q}=\mathcal{\tilde
{Q}}-\mathcal{R}.\label{rel}%
\end{equation}
The coefficient $\mathcal{A}\equiv\partial\Omega_{s}/\partial w$ is given by:%
\begin{align}
\mathcal{A}\left(  w\right)   &  =-\int\frac{d\mathbf{k}}{\left(  2\pi\right)
^{3}}\left(  \frac{1}{2E_{\mathbf{k}}}\frac{\sinh(\beta E_{\mathbf{k}})}%
{\cosh(\beta E_{\mathbf{k}})+\cosh(\beta\zeta)}\right.  \nonumber\\
&  \left.  -\frac{1}{2k^{2}}\right)  -\frac{1}{8\pi a_{s}}.\label{A}%
\end{align}
Note that within the local-density approximation (LDA) time and space
derivatives in (\ref{eqmot}) are neglected, and we arrive at the known gap
equation for a uniform Fermi superfluid%
\begin{equation}
\mathcal{A}\left(  w\right)  =0.\label{LDA}%
\end{equation}

In the BCS-BEC crossover regime, the coefficient $\mathcal{E}$ in
(\ref{eqmot}) is, in general, not negligible. This leads to mixing of $\Psi$
and $\bar{\Psi}$ in the evolution equations. This mixing is not surprising. In
the particular case when space and time variations of the order parameter
about its saddle-point value $\delta\Psi\equiv\Psi-\Delta$ are small, these
variations are equivalent to the Gaussian fluctuations \cite{SadeMeloPRL71}.
For temperatures below $T_{c}$, the fluctuation action is a non-diagonal
quadratic form: it contains terms which mix conjugate and non-conjugate pair
fields \cite{Taylor2}.

Taking the limit $T\rightarrow T_{c}$ in the present approach and neglecting
the second-order time derivatives, we can compare the effective field action
of the present work with the results of the standard Ginzburg - Landau type
theory \cite{SadeMeloPRL71,HuangPRA79}. In taking the $T\rightarrow T_{c}$
limit, we also expand the thermodynamic potential with respect to
$w=\left\vert \Psi\right\vert ^{2}$ up to quartic order in the pair field:
$\Omega_{s}^{GL}=\left.  \Omega_{s}\right\vert _{w=0}-aw+bw^{2}/2$. The
coefficients $a$ and $b$ obtained in this way coincide \emph{exactly} with
those given in Ref. \cite{SadeMeloPRL71}. However, performing the summations
over $p$ in (\ref{Sp}) before taking the limit $w\rightarrow0$ we find that
the coefficient $\mathcal{D}$ differs from the coefficient $d$ of Ref.
\cite{SadeMeloPRL71}:%
\begin{equation}
\lim_{w\rightarrow0}\mathcal{D}\left(  w\right)  =\frac{1}{4}\int
\frac{d\mathbf{k}}{\left(  2\pi\right)  ^{3}}\left(  \frac{\tanh\frac{\beta
\xi_{\mathbf{k}}}{2}}{\xi_{\mathbf{k}}^{2}}-\frac{\beta}{2\xi_{k}\cosh
^{2}\frac{\beta\xi_{\mathbf{k}}}{2}}\right)  . \label{D0}%
\end{equation}
It remains real for all $T<T_{c}$, whereas an imaginary part appears just when
the order of limits $w\rightarrow0$ and $T\rightarrow T_{c}$ is reversed. This
difference is explained by the fact that in Ref. \cite{SadeMeloPRL71},
$\left\vert \Psi\right\vert $ is a small parameter, so that the chemical
potential $\mu$ plays the role of the energy scale. There is a point close to
the unitarity regime where both $\mu$ and $\left\vert \Psi\right\vert $ turn
to zero. In this singularity point, the energy scale vanishes and the time
dependent Ginzburg Landau description (TDGL), as concluded in
\cite{SadeMeloPRL71}, fails. Contrary to the regime near $T_{c}$, a TDGL
equation is obtained in Ref. \cite{SadeMeloPRL71} for all couplings at $T=0$.
In this case a nonzero order parameter \textquotedblleft precludes a vanishing
energy scale, and the low frequency expansion of the effective action is well
defined for all couplings\textquotedblright\ \cite{SadeMeloPRL71}. In the
present approach, $\left\vert \Psi\right\vert $ is \emph{not} a small
parameter, because we performed the summation of the effective field action
over the whole series in powers of $\left\vert \Psi\right\vert $. Therefore
the derived effective field action is valid not only for the zero temperature
case but also for the whole range of temperatures below $T_{c}$, except,
maybe, for a vicinity of the aforesaid singular point.

The coefficient at the first time derivative obtained in the present work is
verified by the comparison with the corresponding coefficient found in Refs.
\cite{Babaev,Botelho2006} for a Fermi gas in 2D keeping only the phase
fluctuations but without assuming the phase to be small. This confirms the
importance of a correct sequence of limits: $T\rightarrow T_{c}$ and
$\left\vert \Psi\right\vert \rightarrow0$. When $\left\vert \Psi\right\vert
=0$ is set from the very beginning, as in Ref. \cite{SadeMeloPRL71}, and then
$T$ varies, a singularity appears at $T=T_{c}$ and $\mu=0$. On the contrary,
the coefficient (\ref{D0}) contains no singularity when passing the point
$\mu=0$.

\subsection{Thermodynamic potential \label{subsec2b}}

As established in Refs. \cite{Abrahams,Gorkov}, for a\ BCS superconductor the
dynamic part of the effective action must be, in general, time-nonlocal and
contain both propagating and dissipative parts. In the weak-coupling BCS
superconductors the propagating part is less than the damped one
\cite{Ebisawa}. However, in the atomic Fermi gases, the propagating component
plays an important role because of the presence of the condensed molecular
bosons whose dynamics is primarily the conserved one \cite{Machida}. The
developed formalism catches the non-dissipative part of the time-dependent
term in the effective action. Thus the evolution equation for the order
parameter (when neglecting the second-order time derivatives) is governed by a
time-dependent nonlinear Schr\"{o}dinger equation \cite{Kim,Manini} (rather
than a time dependent Ginzburg - Landau equation, which must account for the
carrier dissipation).%

\begin{figure}
[h]
\begin{center}
\includegraphics[
height=2.3486in,
width=3.1502in
]%
{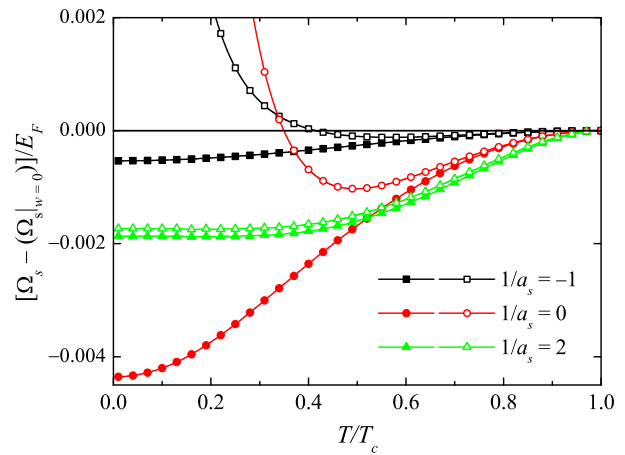}%
\caption{Thermodynamic potential difference $\Omega_{s}-\left(  \left.
\Omega_{s}\right\vert _{w=0}\right)  $ calculated within the finite
temperature EFT (full symbols) and within the standard GL theory
\cite{SadeMeloPRL71} (hollow symbols) as a function of the temperature for
different values of the inverse scattering length.}%
\label{fig:Omega}%
\end{center}
\end{figure}

Figure \ref{fig:Omega} shows the difference $\Omega_{s}-\left(  \left.
\Omega_{s}\right\vert _{w=0}\right)  $ as a function of temperature for
several values of the inverse scattering length $1/a_{s}$, for the present
approach (full curves) and the standard GL approach (dashed curves). According
to the chosen system of units, $a_{s}$ is measured in units of the inverse to
the Fermi wave vector $k_{F}$, and $\Omega_{s}$ is measured in units of
$E_{F}$. The same units are assumed in the other figures. As discussed in the
previous paragraph, near $T_{c}$ the results are close to each other. For
$T\rightarrow0$, the present approach converges to the result for the ground
state energy of the superfluid Fermi gas obtained in the microscopic theory of
the homogeneous system. Indeed, in the limit of a stationary and homogeneous
system without vortices, the minimization of the effective action (\ref{FGL2})
obviously leads to the saddle-point gap equation of Ref. \cite{SadeMeloPRL71}
for all temperatures. In contrast, the standard GL approach is seen to fail
for $T\ll T_{c}$, and does so more strongly for negative scattering lengths.

\subsection{Finite-temperature vortex \label{subsec2c}}

Vortices in superfluid Fermi gases in the BCS-BEC crossover have been studied
with several methods. The vortex core structure was elucidated within a
Bogoliubov - de Gennes approach by Simonucci \emph{et al}.
\cite{Simonucci2013}. The BdG results of Ref. \cite{Simonucci2013} describe an
isolated vortex beyond the weak-coupling BCS case, and in the whole
temperature range $0<T<T_{c}$. Also the present effective field theory allows
us to investigate the vortex core structure at arbitrary temperature and
coupling strengths, and has the advantage of requiring much less computational
effort. Here, we compare the results from the present treatment to the BdG results.

\begin{figure}
[h]
\begin{center}
\includegraphics[
height=4.9896in,
width=2.9307in
]%
{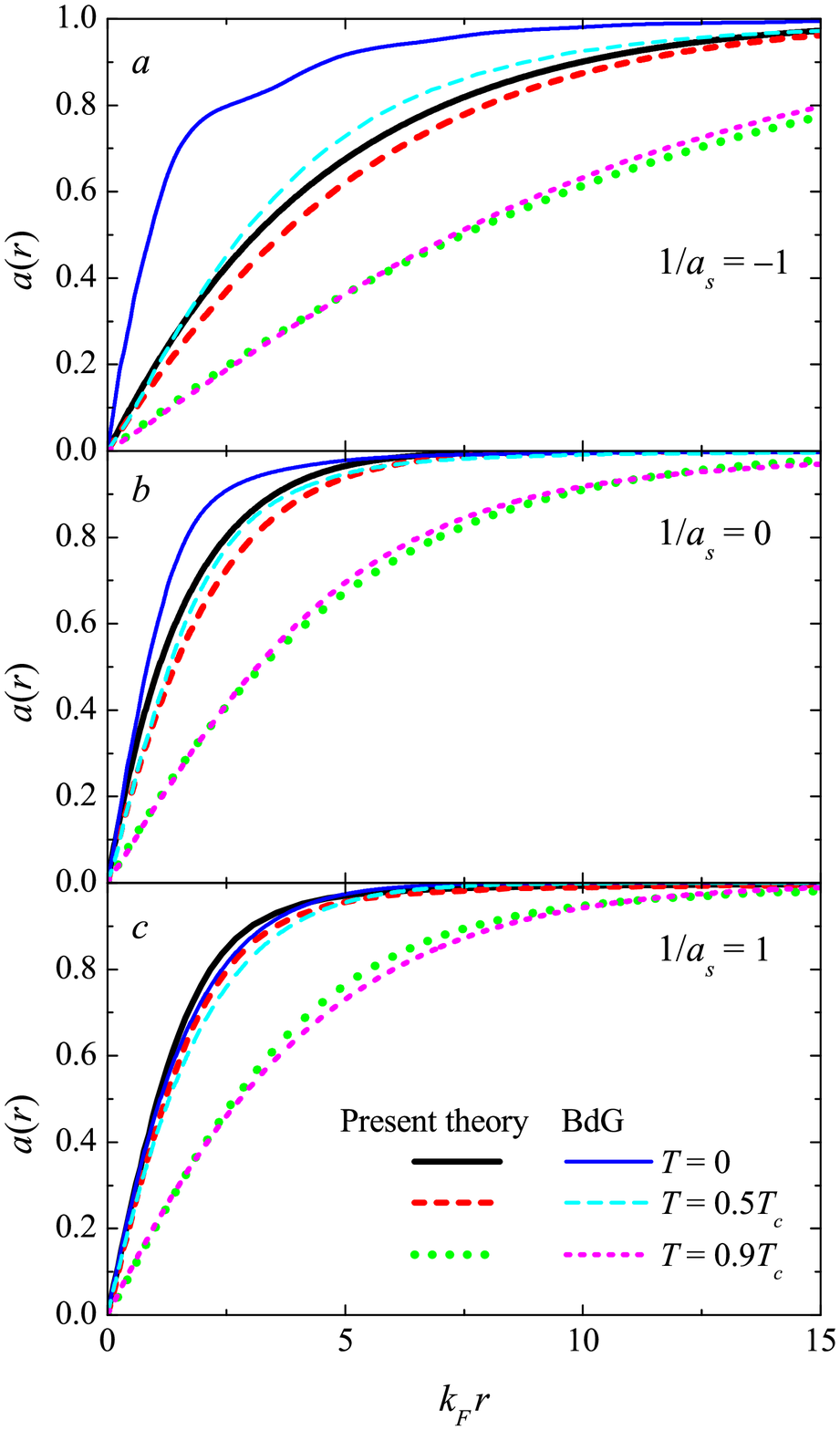}%
\caption{Amplitude modulation function of the order parameter $a\left(
r\right)  =\left\vert \Psi\left(  r\right)  \right\vert /\left\vert
\Psi\left(  \infty\right)  \right\vert $ for a vortex at different
temperatures and scattering lengths. The results of the present theory (heavy
curves) are compared with the BdG data of Ref. \cite{Simonucci2013} (thin
curves).Amplitude modulation function of the order parameter $a\left(
r\right)  =\left\vert \Psi\left(  r\right)  \right\vert /\left\vert
\Psi\left(  \infty\right)  \right\vert $ for a vortex at different
temperatures and scattering lengths. The results of the present theory (heavy
curves) are compared with the BdG data of Ref. \cite{Simonucci2013} (thin
curves).}%
\label{fig:vort1}%
\end{center}
\end{figure}

In Fig.~\ref{fig:vort1}, the amplitude modulation function $a\left(  r\right)
\equiv\left\vert \Psi\left(  r\right)  \right\vert /\left\vert \Psi\left(
\infty\right)  \right\vert $ for a vortex is plotted for several inverse
scattering lengths $1/a_{s}$ and several temperatures. In Ref.
\cite{Simonucci2013}, three temperatures are considered for each scattering
length: $T/T_{c}=0$, $0.5,$ and $0.9$. We use the same temperatures, except
$T=0:$ the low-temperature curves are calculated here for $T/T_{F}=0.005$. At
low temperatures the calculated results very slowly depend on $T$, so that we
can compare our low-temperature results with those for $T=0$ from Ref.
\cite{Simonucci2013}. We find that the agreement is good as the temperature
becomes larger or the interaction regime goes towards the BEC regime. A
significant quantitative difference between BdG and EFT appears only in the
BCS regime at low temperatures.%

%

\begin{figure}
[h]
\begin{center}
\includegraphics[
height=4.9878in,
width=2.9445in
]%
{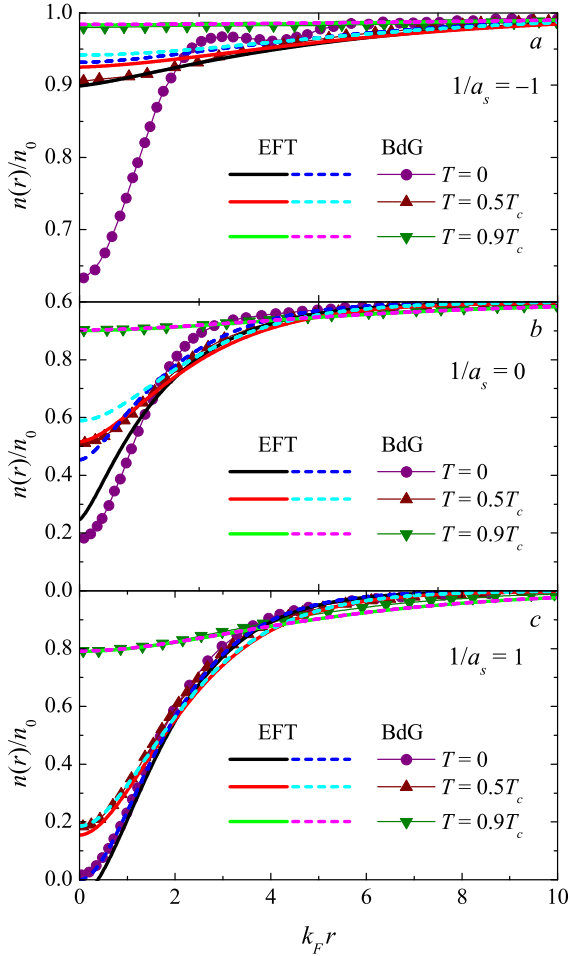}%
\caption{Density distribution (in units of the bulk density $n_{0}$) for a
vortex at different temperatures and scattering lengths. The results of the
present theory (curves) are compared with the BdG data of Ref.
\cite{Simonucci2013} (symbols). The density calculated within LDA is shown by
the solid curves, and the density calculated accounting for the gradient terms
is shown by the dashed curves.}%
\label{fig:vort2}%
\end{center}
\end{figure}

In Fig.~\ref{fig:vort2}, we plot the distributions of the total fermion
density, comparing BdG and EFT results. The density can be calculated by two
methods: (1) in the local density approximation (LDA):%
\begin{equation}
n^{\left(  \text{LDA}\right)  }=-\frac{\partial\Omega_{s}}{\partial\mu},
\label{nLDA}%
\end{equation}
and (2) accounting for the gradient terms in the effective action
(\ref{SEFT}),%
\begin{equation}
n^{\left(  tot\right)  }=-\frac{\partial\Omega_{s}}{\partial\mu}-\frac{1}%
{2}\frac{\partial\rho_{qp}}{\partial\mu}\left(  \frac{da\left(  r\right)
}{dr}\right)  ^{2}-\frac{1}{2r^{2}}\frac{\partial\rho_{sf}}{\partial\mu}%
a^{2}\left(  r\right)  .
\end{equation}
with the superfluid density $\rho_{sf}$ and the quantum pressure coefficient
$\rho_{qp}$:%
\begin{align}
\rho_{sf}  &  =\frac{\mathcal{\tilde{C}}}{m}\left\vert \Psi\right\vert
^{2},\label{rsf}\\
\rho_{qp}  &  =\frac{\left(  \mathcal{\tilde{C}}-4\mathcal{E}\right)
\Delta^{2}}{m}. \label{rqp}%
\end{align}
The superfluid density determined by (\ref{rsf}) explicitly leads to the
expression
\begin{equation}
\rho_{sf}=\frac{\left\vert \Psi\right\vert ^{2}}{3m^{2}}\int\frac{d\mathbf{k}%
}{\left(  2\pi\right)  ^{3}}k^{2}~f_{2}\left(  \beta,E_{\mathbf{k}}%
,\zeta\right)  . \label{rsf1}%
\end{equation}
Remarkably, this expression corresponds exactly to the Landau-type formula for
a Fermi superfluid, but now extended throughout the whole BCS-BEC crossover,
similarly to Ref. \cite{Taylor2006}. The total superfluid density, as shown in
Ref. \cite{Taylor2006}, consists of two parts: the mean-field contribution,
that is equivalent to (\ref{rsf1}), and a fluctuation contribution. The
fluctuation contribution was also considered in the microscopic rederivations
of the Berezinskii-Kosterlitz-Thouless theory based on a path-integral
treatment of phase fluctuations in two-dimensional Fermi gases within the
low-wavelength approximation \cite{Babaev,Botelho2006,BKT-PRA2009}. The
superfluid density entering the phase action as a prefactor at $\left(
\nabla\theta\right)  ^{2}$ in these works can be obtained from (\ref{rsf1}) by
a straightforward translation of the present formalism to the two-dimensional
case. These two examples represent a reassuring analytic verification of the
present formalism as they agree with well-established preceding results.

As was also the case for the order parameter, the agreement between BdG and
EFT is gradually better for higher temperatures and/or when moving to the BEC
side, where EFT retrieves the Gross-Pitaevskii theory. The gradient
corrections improve the agreement between BdG and EFT in the BCS and unitarity
regimes. However, in the BEC regime the gradient corrections are extremely
small, except at $T=0$. In the low-temperature limit, the gradient corrections
in the BEC regime result in a small artifact: the density goes to negative
values near the vortex center. Thus in the BEC regime, LDA seems to describe
the density better than the calculation including the gradient corrections.
These results are in agreement with the recent work of Ref.
\cite{Simonucci2014} (citing our approach in Ref. [16] of that paper) where a
long-wavelength approximation has been developed for the BdG equations. That
approach differs from the present formalism in that we perform the
long-wavelength expansion for the exact effective bosonic action rather than
for the BdG equations (which are already an approximation). Nevertheless, the
results of these two approaches are close to each other.

\subsection{Collective excitations \label{subsec2d}}

The spectrum of the collective excitations is determined in the following way,
similarly to Ref. \cite{Diener2008}. First, we assume that the pair field
$\Psi$ is a sum of the uniform and time-independent mean-field value $\Delta$
and the fluctuation field $\varphi$:%
\begin{equation}
\Psi\left(  \mathbf{r},\tau\right)  =\Delta+\varphi\left(  \mathbf{r}%
,\tau\right)  ,\quad\bar{\Psi}\left(  \mathbf{r},\tau\right)  =\Delta
+\bar{\varphi}\left(  \mathbf{r},\tau\right)  \label{fluct}%
\end{equation}
and keep the fluctuations up to second order. Next, the pair field is
rewritten in the $\left(  q,i\Omega_{n}\right)  $ representation. This gives
us the quadratic fluctuation action in matrix form:%
\begin{align}
S_{EFT}^{\left(  quad\right)  }  &  =\frac{1}{2}\sum_{\mathbf{q},n}\left(
\begin{array}
[c]{cc}%
\bar{\varphi}_{\mathbf{q},n} & \varphi_{-\mathbf{q},-n}%
\end{array}
\right) \nonumber\\
&  \times\mathbb{M}\left(  q,i\Omega_{n}\right)  \left(
\begin{array}
[c]{c}%
\varphi_{\mathbf{q},n}\\
\bar{\varphi}_{-\mathbf{q},-n}%
\end{array}
\right)  , \label{Squad}%
\end{align}
where the matrix $\mathbb{M}\left(  q,i\Omega_{n}\right)  $ is determined by:%
\begin{align}
M_{1,1}\left(  q,i\Omega_{n}\right)   &  =\mathcal{U}+\frac{\mathcal{C}}%
{2m}q^{2}-i\Omega_{n}\mathcal{\tilde{D}}+\Omega_{n}^{2}\mathcal{Q},\nonumber\\
M_{1,2}\left(  q,i\Omega_{n}\right)   &  =\mathcal{U}-\frac{\mathcal{E}}%
{m}q^{2}-\mathcal{R}\Omega_{n}^{2},\nonumber\\
M_{2,1}\left(  q,i\Omega_{n}\right)   &  =M_{1,2}\left(  q,-i\Omega
_{n}\right)  ,\nonumber\\
M_{2,2}\left(  q,i\Omega_{n}\right)   &  =M_{1,1}\left(  q,-i\Omega
_{n}\right)  , \label{M}%
\end{align}
with the coefficients introduced in (\ref{Ws})--(\ref{rr}) and%
\begin{align}
\mathcal{U}\left(  w\right)   &  =w\frac{\partial^{2}\Omega_{s}\left(
w\right)  }{\partial w^{2}},\quad\mathcal{\tilde{D}}\left(  w\right)
=\frac{\partial\left[  w\mathcal{D}\left(  w\right)  \right]  }{\partial
w},\label{coefs}\\
\mathcal{C}  &  =\mathcal{\tilde{C}}-2\mathcal{E},\quad\mathcal{Q}%
=\mathcal{\tilde{Q}}-\mathcal{R}.
\end{align}

The spectra of collective excitations are determined after the transition
$i\Omega_{n}\rightarrow\omega$ as the roots of the equation%
\begin{equation}
\det\mathbb{M}\left(  q,\omega\right)  =0. \label{det}%
\end{equation}
The solution of equation (\ref{det}) in the long-wavelength approximation
yields the Bogoliubov -- Anderson (Goldstone) mode with the frequency%
\begin{equation}
\omega_{q}=v_{s}q, \label{wq}%
\end{equation}
where $v_{s}$ is the first sound velocity. It is expressed through the
coefficients of the effective field action similarly to Refs.
\cite{Diener2008,NJP2012}:%
\begin{equation}
v_{s}=\sqrt{\frac{1}{m}\frac{\mathcal{U\tilde{C}}}{\mathcal{\tilde{D}}%
^{2}+2\mathcal{U\tilde{Q}}}}. \label{vs}%
\end{equation}
Note that the coefficient $\mathcal{\tilde{Q}}$ corresponding to the second
order of the derivative expansion for the time derivatives enters the sound
velocity together with the first order coefficient $\mathcal{\tilde{D}}$. This
result demonstrates that the second order of the derivative expansion is
important for the spectrum of collective excitations.

In the zero temperature limit, the coefficients entering the matrix (\ref{M})
for the quadratic fluctuation action correspond exactly to those obtained
within the Gaussian pair fluctuation (GPF) theory \cite{Diener2008} and in the
zero-temperature theory of Ref. \cite{Marini1998}. Remarkably, despite the
fact that the approach of Ref. \cite{Marini1998} is non-perturbative (i.e.
without assuming the non-uniform part of the pair field to be small), the
coefficients for the zero-temperature action functional in Ref.
\cite{Marini1998} appear to be the same as in the GPF at $T=0$
\cite{Diener2008}. In other words, at zero temperature, the result of two
approximations (small fluctuations and slowly varying fluctuations) does not
depend on their sequence. On the contrary, at nonzero temperatures these two
approximations do not commute.%

\begin{figure}
[h]
\begin{center}
\includegraphics[
height=2.3827in,
width=3.0617in
]%
{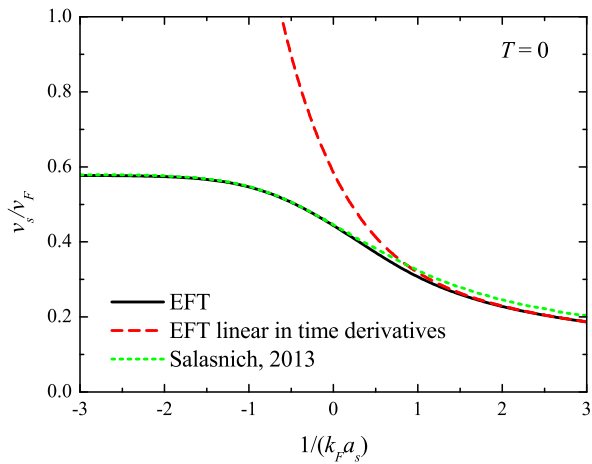}%
\caption{The sound velocity $v_{s}$ calculated using formula (\ref{vs}) (solid
curve) and neglecting the second time derivative (dashed curve), compared with
the result of Ref. \cite{Salasnich2013} for a 3D Fermi gas.}%
\label{fig:vs1}%
\end{center}
\end{figure}

In Fig.~\ref{fig:vs1}, the sound velocity $v_{s}$ (in units of the Fermi
velocity $v_{F}\equiv\hbar k_{F}/m$) calculated using the mean-field values of
the chemical potential is plotted as a function of the inverse scattering
length and compared with that extracted from Ref. \cite{Salasnich2013} for a
Fermi gas in three dimensions. In that paper, the effect of both phase and
amplitude fluctuations of the order parameter is taken into account in the
determination of the sound velocity of the uniform superfluid system in the
BCS-BEC crossover. The results obtained in Ref. \cite{Salasnich2013} depend
strongly on whether amplitude fluctuations are taken into account or not. The
amplitude fluctuations are incorporated in Ref. \cite{Salasnich2013} following
Schakel \cite{Schakel}, obtaining results at the unitarity regime and at
$T=0$. The present calculation also takes into account both phase and
amplitude fluctuations, at all coupling strengths and temperatures. We can see
from Fig.~\ref{fig:vs1} that our result for $v_{s}$ and that of Ref.
\cite{Salasnich2013} agree excellently at the BCS side, and exhibit only a
slight difference in the BEC regime.

In order to show the importance of the terms of second order in the time
derivative, we show in Fig.~\ref{fig:vs1} also the sound velocity determined
within our EFT neglecting the coefficient $\mathcal{\tilde{Q}}$ in (\ref{vs}).
It is clear that setting $\mathcal{\tilde{Q}}=0$ leads to a substantial change
in the BCS regime, while leaving the result in the BEC regime unaffected.

\section{Two-band Fermi superfluids \label{TwoBand}}

\subsection{Extension of the EFT to two bands \label{subsec3a}}

The extension of the EFT formulated in the above subsection to the two-band
Fermi systems is particularly interesting due to recent intense discussions on
the applicability of the GL approach to the coupled Fermi systems far below
$T_{c}$
\cite{KoganPRB83,KoganPRB86,BabaevPRB86,ShanenkoPRL106,Shanenko2,Babaev2011,Babaev2012-2}%
.

Here, we consider a fermionic system of two types of particles ($j=1,2$) with
two spin states each described by the microscopic atomic Hamiltonians
$H_{\sigma,j}$ as in the section on the one-band system, with possibly
different masses and chemical potentials for each band. The path-integral
scheme remains the same as above. The interaction Hamiltonian $U\left(
\mathbf{r},\tau\right)  $, however, is more complicated, because it describes
both intraband and interband interactions:%
\begin{align}
U  &  =%
{\textstyle\sum\nolimits_{j=1,2}}
g_{j}\bar{\psi}_{\uparrow,j}\bar{\psi}_{\downarrow,j}\psi_{\downarrow,j}%
\psi_{\uparrow,j}\nonumber\\
&  +g_{3}\left(  \bar{\psi}_{\uparrow,1}\psi_{\uparrow,1}\bar{\psi
}_{\downarrow,2}\psi_{\downarrow,2}+\bar{\psi}_{\downarrow,1}\psi
_{\downarrow,1}\bar{\psi}_{\uparrow,2}\psi_{\uparrow,2}\right) \nonumber\\
&  +g_{4}\left(  \bar{\psi}_{\uparrow,1}\psi_{\uparrow,1}\bar{\psi}%
_{\uparrow,2}\psi_{\uparrow,2}+\bar{\psi}_{\downarrow,1}\psi_{\downarrow
,1}\bar{\psi}_{\downarrow,2}\psi_{\downarrow,2}\right)  . \label{U}%
\end{align}

The terms with the coupling constants $g_{1},g_{2}$ determine the intraband
scattering between two fermions of the same type and with antiparallel spins:
these are the two Cooper pairing channels. The terms with $g_{3}$ and $g_{4}$
are related to the interband scattering for the fermions with antiparallel and
parallel spins, respectively. In ultracold gases, scattering between fermions
with parallel spins is not present for fermions in the same band due to the
Pauli principle. However, parallel spin scattering between fermions in
\emph{different} bands should be kept, as it contributes to the
renormalization of the effective interaction. Terms of the type $\bar{\psi
}_{\downarrow,1}\bar{\psi}_{\uparrow,1}\psi_{\downarrow,2}\psi_{\uparrow,2}$
are not included in the interaction Hamiltonian (\ref{U}). They are kept in
some theoretical schemes (e. g. \cite{ZhitomirskyPRB69,Iskin,Iskin2}) and
describe an \emph{ad hoc} model interband scattering of pairs. However, we
avoid such terms in the starting microscopic action because they cannot arise
from any density-density type of interaction.

The Hubbard-Stratonovich (HS) transformation is based on introducing auxiliary
fields $\Psi_{j}$ and $\chi_{j}$ such that the relation%
\begin{equation}
\mathcal{Z}\propto\int\mathcal{D}\left[  \bar{\psi},\psi\right]
\int\mathcal{D}\left[  \bar{\Psi},\Psi\right]  \int\mathcal{D}\left[
\bar{\chi},\chi\right]  e^{-S_{HS}} \label{Z1}%
\end{equation}
is satisfied. In the HS action $S_{HS}$, the fermion fields appear only up to
quadratic order so they can be integrated. The HS action which exactly
decouples the four-field interaction terms in the initial Hamiltonian,
involves two pair fields and two density fields corresponding to the interband
normal channel (see Ref. \cite{Kleinert}):%
\begin{align}
S_{HS}  &  =S_{0}+S_{B}+S_{\chi}\nonumber\\
&  +\sum_{j=1,2}\int_{0}^{\beta}d\tau\int d\mathbf{r}\left(  \bar{\Psi}%
_{j}\psi_{j,\uparrow}\psi_{j,\downarrow}+\Psi_{j}\bar{\psi}_{j,\downarrow}%
\bar{\psi}_{j,\uparrow}\right. \nonumber\\
&  \left.  +i\bar{\chi}_{j}\rho_{j}+i\chi_{j}\bar{\rho}_{j}\right)  ,
\label{SHS}%
\end{align}
where $\rho_{1}=\bar{\psi}_{1,\uparrow}\psi_{2,\downarrow}+\bar{\psi
}_{2,\uparrow}\psi_{1,\downarrow}$ and $\rho_{2}=\bar{\psi}_{1,\uparrow}%
\psi_{2,\uparrow}+\bar{\psi}_{2,\downarrow}\psi_{1,\downarrow}$ are
combinations of the fermion variables, $\Psi_{j}$ and $\chi_{j}$ are the HS
pair and density fields, respectively. The actions of the free HS fields are
given by:%
\begin{align}
S_{B}  &  =-\int_{0}^{\beta}d\tau\int d\mathbf{r}\left[  \frac{1}{G_{1}}%
\bar{\Psi}_{1}\Psi_{1}+\frac{1}{G_{2}}\bar{\Psi}_{2}\Psi_{2}\right.
\nonumber\\
&  \left.  -\frac{1}{G_{12}}\left(  \bar{\Psi}_{1}\Psi_{2}+\bar{\Psi}_{2}%
\Psi_{1}\right)  \right]  ,\label{SB}\\
S_{\chi}  &  =-\int_{0}^{\beta}d\tau\int d\mathbf{r}\left(  \frac{1}{g_{3}%
}\bar{\chi}_{1}\chi_{1}+\frac{1}{g_{4}}\bar{\chi}_{2}\chi_{2}\right)  .
\label{Sxi}%
\end{align}
The intraband channel for same-spin fermions is not present (nor is it in Ref.
\cite{SadeMeloPRL71}) because we assume the temperature is low enough so that
only $s$-wave scattering occurs. The four-field HS transformation exactly
eliminates the fermion-fermion interaction from the initial Hamiltonian. If
the interband coupling is switched off, the effective bosonic action exactly
turns to that exploited in Ref. \cite{SadeMeloPRL71} for two independent bands.

Although there is no Josephson interband coupling in the initial
fermion-fermion interaction (\ref{U}), this coupling emerges in a natural way
in the effective bosonic action (\ref{SB}) and follows explicitly from the HS
transformation of the microscopic action. The coupling constants $G_{j}$ are
related to those from (\ref{U}) in the following way:%
\begin{align}
\frac{1}{G_{1}}  &  =\frac{g_{2}}{g_{1}g_{2}-g_{12}^{2}},\;\frac{1}{G_{2}%
}=\frac{g_{1}}{g_{1}g_{2}-g_{12}^{2}},\nonumber\\
\frac{1}{G_{12}}  &  =\frac{g_{12}}{g_{1}g_{2}-g_{12}^{2}},\;g_{12}%
=g_{4}-g_{3}. \label{CC}%
\end{align}

In order to address the whole range of the BCS-BEC crossover, the coupling
constants $g_{1},g_{2}$ are renormalized through the $s$-wave scattering
lengths $a_{s,j}$ similarly to Ref. \cite{SadeMeloPRL71} and in the above
subsection for the one-band system:
\begin{equation}
\frac{1}{g_{j}}=m_{j}\left(  \frac{1}{4\pi a_{s,j}}-\int_{k<K}\frac
{d\mathbf{k}}{\left(  2\pi\right)  ^{3}}\frac{1}{k^{2}}\right)  .
\label{renorm1}%
\end{equation}
with $K\rightarrow\infty$. In order to ensure convergence in the thermodynamic
potential and in the gap equation, the other two coupling constants
$g_{3},g_{4}$ must also be renormalized through Eq. (\ref{renorm1}), with the
scattering lengths $a_{s,3}$ and $a_{s,4}$ and the mass parameter $m_{3}%
=m_{4}\equiv m_{12}$. The mass parameter $m_{12}$, as shown below, enters the
final results through the factor $\gamma m_{12}$ with the interband coupling
parameter $\gamma\equiv2\left(  \frac{1}{a_{s,3}}-\frac{1}{a_{s,4}}\right)  $.
Consequently this mass can be chosen chosen arbitrary as far as the
renormalization is concerned, and we set $m_{12}=\sqrt{m_{1}m_{2}}$. With
these renormalizations, $g_{12}^{2}/\left(  g_{1}g_{2}\right)  \propto1/K^{2}%
$, so that the stability condition $g_{1}g_{2}>g_{12}^{2}$ is always fulfilled.

The integration over the fermion fields leads to the partition function,%
\begin{equation}
\mathcal{Z}\propto\int\mathcal{D}\left[  \bar{\Psi},\Psi\right]
\int\mathcal{D}\left[  \bar{\chi},\chi\right]  e^{-S_{eff}}, \label{Z2a}%
\end{equation}
with the effective bosonic action $S_{eff}$. The details for the effective
bosonic action are described in the Appendix. The ultraviolet-divergent part
of the effective bosonic action can be explicitly extracted. When introducing
a sufficiently large momentum cutoff $k_{0}$ for the fermion fields, the part
of the HS action for $k>k_{0}$ provides the ultraviolet-divergent part of the
effective bosonic action%
\begin{align}
\delta S_{eff}^{\left(  \operatorname{div}\right)  }\left(  k_{0}\right)   &
=-\int_{0}^{\beta}d\tau\int d\mathbf{r}\int_{\left(  k>k_{0}\right)  }%
\frac{d\mathbf{k}}{\left(  2\pi\right)  ^{3}}\sum_{j=1,2}\frac{m_{j}}{k^{2}%
}\nonumber\\
&  \times\bar{\Psi}_{j}\left(  \mathbf{r},\tau\right)  \Psi_{j}\left(
\mathbf{r},\tau\right)  +O\left(  k^{-4}\right)  . \label{div2}%
\end{align}
The density fields $\chi_{j}$ do not contribute to the ultraviolet divergence
of the effective bosonic action. In the limit $\left\vert g_{j}\right\vert
\rightarrow\infty$ (corresponding to $K\rightarrow\infty$)$,$ the divergence
of the action $S_{B}$ is exactly compensated by (\ref{div2}) using
(\ref{renorm1}), so that the part of the effective action depending on the
pair fields $\Psi_{j}$ is regularized. On the contrary, the density-field
action $S_{\chi}$ unrestrictedly increases when $K\rightarrow\infty$. Thus the
functional $e^{-S_{\chi}}$ acts as a product of delta functions for the
density fields $\chi_{j}\left(  \mathbf{r},\tau\right)  $ at all $\left(
\mathbf{r},\tau\right)  $. As a result, the subsequent integration over the
density fields is performed exactly, and we arrive at the effective bosonic
action depending on the pair fields only:%
\begin{equation}
S_{eff}=S_{B}-\sum_{j=1,2}\operatorname{Tr}\ln\left[  -\mathbb{G}_{j}%
^{-1}\right]  . \label{Seff}%
\end{equation}
The inverse Nambu tensor $\mathbb{G}_{j}^{-1}$ for each band has been
determined above. In the effective field action (\ref{Seff}), the coupling
strengths of the starting Hamiltonian are reduced to only three input
parameter of the theory: two scattering lengths $a_{s,1},a_{s,2}$ and the
Josephson interband coupling strength $\gamma$. The intraband coupling
strengths (expressed here through the scattering lengths) and the interband
coupling strength are the standard input parameter which are used in known
works of two-band superconductivity/superfluidity, see, e. g. Refs.
\cite{SuhlPRL3,Iskin,Iskin2,Iskin3,ZhitomirskyPRB69}. These parameter are
measurable: the scattering lengths and the Josephson interband coupling
strength can be experimentally determined and controlled using e. g. the
Feshbach resonance.

The effective action (\ref{Seff}) is expanded as a series in powers of the
pair field in the same way as for a one-band system. Correspondingly, the
derivative expansion and the summation over the whole series in powers of the
pair fields is performed for each band independently. As a result, the bosonic
action $S_{eff}$ is approximated by the following effective field action
$S_{EFT}^{\left(  2b\right)  }$ :%
\begin{equation}
S_{EFT}^{\left(  2b\right)  }=\sum_{j=1,2}S_{EFT}^{\left(  j\right)  }%
-\int_{0}^{\beta}d\tau\int d\mathbf{r~}\frac{m_{12}\gamma}{4\pi}\left(
\bar{\Psi}_{1}\Psi_{2}+\bar{\Psi}_{2}\Psi_{1}\right)  , \label{SEFT}%
\end{equation}
where $S_{EFT}^{\left(  j\right)  }$ is the effective field action for each
band determined by (\ref{FGL2}).

The equations of motion for the order parameters follow from the stationary
action principle, resulting in a set of four coupled equations. Two equations
-- for $j=1$ and $j=2$ -- are:%
\begin{align}
&  i\frac{\partial\left(  w_{j}\mathcal{D}_{j}\right)  }{\partial w_{j}}%
\frac{\partial\Psi_{j}}{\partial t}=\mathcal{A}_{j}\left(  w_{j}\right)
\Psi_{j}-\frac{m_{12}\gamma}{4\pi}\Psi_{3-j}\nonumber\\
&  +\mathcal{Q}_{j}\frac{\partial^{2}\Psi_{j}}{\partial t^{2}}-\frac
{\mathcal{R}_{j}\Psi_{j}^{2}}{w_{j}}\frac{\partial^{2}\bar{\Psi}_{j}}{\partial
t^{2}}\nonumber\\
&  -\frac{1}{w_{j}}\frac{\partial\left(  w_{j}\mathcal{R}_{j}\right)
}{\partial w_{j}}\Psi_{j}\frac{\partial\bar{\Psi}_{j}}{\partial t}%
\frac{\partial\Psi_{j}}{\partial t}\nonumber\\
&  +\left(  \frac{\partial\mathcal{Q}_{j}}{\partial w_{j}}+\frac{1}{2w_{j}%
}\frac{\partial\left(  w_{j}\mathcal{R}_{j}\right)  }{\partial w_{j}}\right)
\bar{\Psi}_{j}\left(  \frac{\partial\Psi_{j}}{\partial t}\right)
^{2}\nonumber\\
&  -\frac{1}{2}\frac{\partial\left(  \frac{\mathcal{R}_{j}}{w_{j}}\right)
}{\partial w_{j}}\Psi_{j}^{3}\left(  \frac{\partial\bar{\Psi}_{j}}{\partial
t}\right)  ^{2}\nonumber\\
&  -\frac{\mathcal{C}_{j}}{2m_{j}}\nabla_{\mathbf{r}}^{2}\Psi_{j}%
+\frac{\mathcal{E}_{j}\Psi_{j}^{2}}{m_{j}w_{j}}\nabla_{\mathbf{r}}^{2}%
\bar{\Psi}_{j}\nonumber\\
&  +\frac{1}{m_{j}w_{j}}\frac{\partial\left(  w_{j}\mathcal{E}_{j}\right)
}{\partial w_{j}}\Psi_{j}\left(  \nabla_{\mathbf{r}}\bar{\Psi}_{j}\cdot
\nabla_{\mathbf{r}}\Psi_{j}\right) \nonumber\\
&  -\frac{1}{2m_{j}}\left(  \frac{\partial\mathcal{C}_{j}}{\partial w_{j}%
}+\frac{1}{w_{j}}\frac{\partial\left(  w_{j}\mathcal{E}_{j}\right)  }{\partial
w_{j}}\right)  \bar{\Psi}_{j}\left(  \nabla_{\mathbf{r}}\Psi_{j}\right)
^{2}\nonumber\\
&  +\frac{1}{2m_{j}}\frac{\partial\left(  \frac{\mathcal{E}_{j}}{w_{j}%
}\right)  }{\partial w_{j}}\Psi_{j}^{3}\left(  \nabla_{\mathbf{r}}\bar{\Psi
}_{j}\right)  ^{2}, \label{eqmot1}%
\end{align}
and the other two equations are conjugate to (\ref{eqmot1}). Within the
local-density approximation (LDA) time and space derivatives in (\ref{eqmot1})
are neglected and we obtain two coupled gap equations,
\begin{equation}
\mathcal{A}_{j}\left(  w_{j}\right)  \Psi_{j}-\frac{m_{12}\gamma}{4\pi}%
\Psi_{3-j}=0. \label{LDA2b}%
\end{equation}

The two-band effective action allows for mass imbalance $m_{1}\neq m_{2}$.
This can be important for two-band superconductors but seems to be less
relevant for ultracold atomic gases, because phase coherence and Josephson
coupling between species with different masses are hardly achievable. In the
treatment of applications of the EFT in the present paper, we assume no mass
imbalance. Consequently, for the numerical calculations we use the same set of
units as in Sect. \ref{EFT}.

\subsection{Uniform two-band superfluid \label{subsec3b}}

\subsubsection{Parameters and thermodynamic functions at $T=0$}

In two-bandgap Fermi superfluids, the interband interactions compete with the
intraband interactions and affect the ground-state phases. The thermodynamic
potential per unit volume for the uniform two-band system resulting from the
action functional (\ref{SEFT}) is:%
\begin{equation}
\Omega=\sum_{j=1,2}\Omega_{s,j}-\frac{m_{12}\gamma}{4\pi}\left(  \bar{\Psi
}_{1}\Psi_{2}+\bar{\Psi}_{2}\Psi_{1}\right)  . \label{W}%
\end{equation}
At zero temperature, the mean-field thermodynamic potential adequately
describes the properties of the Fermi superfluid in the whole range of the
BCS-BEC crossover. The internal energy as a function of the total fermion
particle density $n$, is determined at $T=0$ through the thermodynamic
relation%
\begin{equation}
E=\Omega+\mu n. \label{E}%
\end{equation}
In the present treatment we assume that the masses of the fermions in the two
subbands are the same, the band offset is equal to zero, and the subbands are
in thermodynamical equilibrium in the sense that their chemical potentials are
equal. The number equation
\begin{equation}
-\frac{\partial\Omega}{\partial\mu}=n. \label{NumEq}%
\end{equation}
determines the chemical potential that is common for both bands. The three
parameters $\Psi_{1},\Psi_{2},\mu$ are found by solving the number equation
(\ref{NumEq}) along with the two coupled gap equations (\ref{LDA2b}). The
roots of this coupled set of equations are derived numerically, and we
investigate the dependence of these solutions on the interband coupling
$\gamma$ and on the intraband coupling parameters $1/a_{s,j}$ where $a_{s,j}$
is the scattering length between fermions in band $j$.%

\begin{figure}
[h]
\begin{center}
\includegraphics[
height=3.8135in,
width=2.7508in
]%
{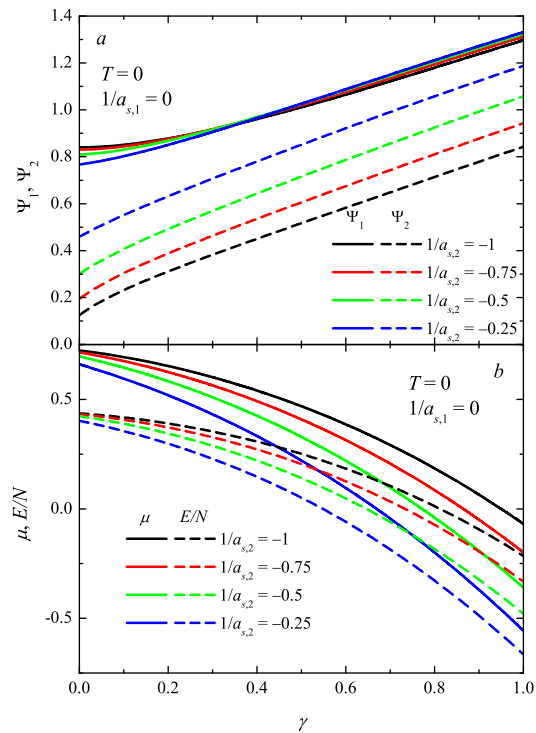}%
\caption{(\emph{a}) Order parameters $\Psi_{1}$ (solid curves) and $\Psi_{2}$
(dashed curves); (\emph{b}) chemical potential (solid curves) and internal
energy (dashed curves) for a two-band superfluid Fermi gas as a function of
the interband coupling strength $\gamma$ for the inverse scattering length of
the \textquotedblleft stronger\textquotedblright\ band $1/a_{1}=0$ and for
different values of $1/a_{2}$.}%
\label{fig:ZT1}%
\end{center}
\end{figure}

In Fig.~\ref{fig:ZT1}~(\emph{a}), the order parameters for a two-band
superfluid Fermi gas are plotted as a function of the interband coupling
strength $\gamma$ for the inverse scattering length of the \textquotedblleft
stronger\textquotedblright\ band $1/a_{s,1}=0$ and for different values of the
inverse scattering length of the \textquotedblleft weaker\textquotedblright%
\ band $1/a_{s,2}$. As intuitively expected, both $\Psi_{1}$ and $\Psi_{2}$
are monotonously increasing functions of the interband coupling strength
$\gamma$. The difference $\Psi_{2}-\Psi_{1}$ is an increasing function of the
difference $1/a_{s,1}-1/a_{s,2}$.

Figure~\ref{fig:ZT1} reveals some surprising details of the interplay between
the interband and intraband couplings. At large $\gamma$, the dependence of
$\Psi_{1}$ on $1/a_{s,2}$ has the same sign as that of $\Psi_{2}$ on
$1/a_{s,2}$. However, at small $\gamma$ (here, $\gamma\lessapprox0.35$), the
order parameter $\Psi_{1}$ becomes a decreasing function of $1/a_{s,2}$ and
this behavior persists even at zero coupling where one expects $\Psi_{1}$ to
be independent of $a_{s,2}$. This behavior of the order parameter can be
explained by a population transfer between bands. Indeed, even in the limit of
the zero coupling, the common chemical potential leads to unequal fermion
densities in the two bands, and in particular, to a depletion of the weak
band.\ The common chemical potential will be affected by the scattering
lengths of both bands, even in the limit of zero interaction coupling between
different bands, and this results in a population transfer between the bands.
The \textquotedblleft stronger\textquotedblright\ band drains away more
fermions from the \textquotedblleft weaker\textquotedblright\ band as the
difference between the inverse scattering lengths $1/a_{s,1}-1/a_{s,2}$ grows.
These additional fermions allow to form more pairs, resulting in an increase
of $\Psi_{1}$ as $1/a_{s,2}$ becomes more negative.

In Fig.~\ref{fig:ZT1}~(\emph{b}), we show the common chemical potential $\mu$
and the internal energy per particle $E/n$ as a function of the interband
coupling strength $\gamma$, for several values of the inverse scattering
$1/a_{s,2}$ length of the \textquotedblleft weaker\textquotedblright\ band. We
can see that an increasing \textit{inter}band coupling acts like an increasing
\textit{intra}band coupling: both the chemical potential and the internal
energy decrease when $\gamma$ rises. In the BEC limit, the difference between
$E/n$ and $\mu$ gradually decreases, while at all coupling $E/n<\mu$. This
inequality must necessarily be fulfilled, because the thermodynamic potential
per unit volume is equal to minus the pressure: $\Omega=-P$ and the pressure
remains positive.

\begin{figure}
[h]
\begin{center}
\includegraphics[
height=4.664in,
width=2.8403in
]%
{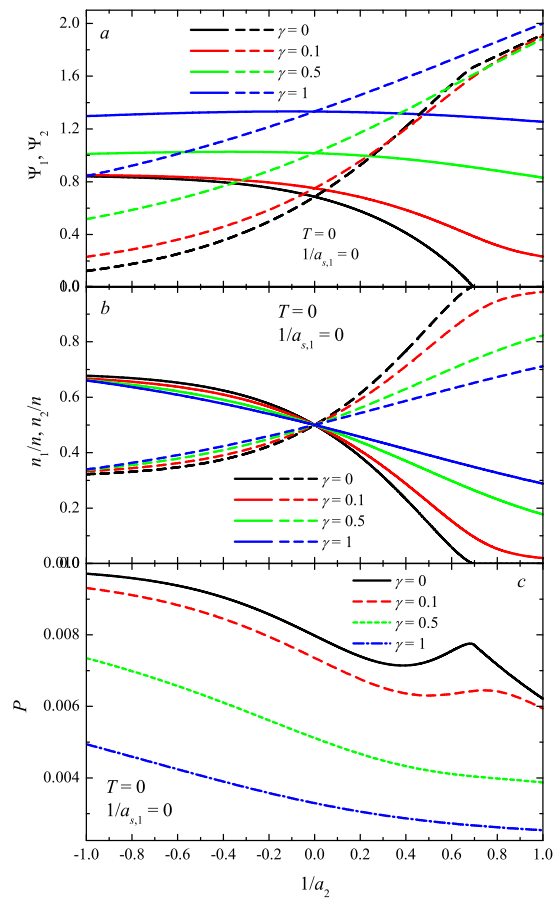}%
\caption{(\emph{a}) Order parameters $\Psi_{1}$ (solid curves) and $\Psi_{2}$
(dashed curves); (\emph{b}) fermion densities in the first band (solid curves)
and second band (dashed curves); (\emph{c}) pressure for a two-band superfluid
Fermi gas as a function of the inverse scattering length $1/a_{s,2}$, for
$1/a_{s,1}=0$ and for different values of the interband coupling strength
$\gamma$.}%
\label{fig:ZT2}%
\end{center}
\end{figure}

In Fig.~\ref{fig:ZT2}~(\emph{a}) we return to the question of population
transfer between the bands, and plot the order parameters as a function of
$1/a_{s,2}$, for $1/a_{s,1}=0$ and for several values of $\gamma$. Here, the
interaction parameter of the second band crosses over from the BCS regime at
$1/a_{s,2}=-1$ to the BEC regime at $1/a_{s,2}=1$. In the range $1/a_{s,2}<0$,
the first band is the \textquotedblleft stronger\textquotedblright\ band, and
for $1/a_{s,2}>0$, the first band is the \textquotedblleft
weaker\textquotedblright\ band. In the BEC regime, we find a remarkable
feature: $\Psi_{1}$ can turn to zero when $1/a_{s,2}$ is sufficiently large
and $\gamma$ is sufficiently small (here, at $\gamma=0$). At this critical
point, also $\Psi_{2}$ reveals a kink.%

The origin of this suppression of $\Psi_{1}$ becomes clear when we look at the
dependence of the relative fermion densities for the first and second band as
a function of $1/a_{s,2}$, shown in Fig.~\ref{fig:ZT2}~(\emph{b}). We again
see that, even when $\gamma=0$, the common chemical potential couples the
bands through population transfer. As $\gamma$ grows, this population transfer
is reduced. In the BEC range, the population transfer is more pronounced, and
when $\gamma$ is small and $1/a_{s,2}$ is sufficiently large, the population
transfer can completely deplete the first band, driving $\Psi_{1}$ down to
zero. This possibility has gone hitherto unnoticed in the theoretical
descriptions of two-band systems, as these are usually studied in the
BCS\ limit. Even in strong-coupling superconductors, the BEC regime seems not
to be achieved. Therefore this effect seems to be a new feature related to the
Fermi gases where the BCS-BEC crossover regime can be experimentally probed.

The effect of band population transfer can also been seen in the dependence of
the pressure $P=-\Omega$ on $1/a_{s,2},$ shown in \ref{fig:ZT2}~(\emph{c}).
The pressure exhibits a local maximum for the curve with $\gamma=0$ when the
\textquotedblleft weaker\textquotedblright\ band becomes completely depleted.
Moreover, when $\gamma$ is nonzero but sufficiently small, the pressure
depends on $1/a_{s,2}$ non-monotonically.

\subsubsection{Temperature dependence of parameters}

As shown in Ref. \cite{Babaev2012-2}, a Josephson coupling for two-bandgap
superconductors yields the symmetry breakdown from $U\left(  1\right)  \times
U\left(  1\right)  $ to $U\left(  1\right)  $ and hence eliminates the
superconducting phase transition for a \textquotedblleft
weaker\textquotedblright\ band at $T_{c,2}$, where $T_{c,2}$ is the critical
temperature for the \textquotedblleft weaker\textquotedblright\ band in the
absence of interband coupling. As a result, the divergence of the coherence
length is removed for the \textquotedblleft weaker\textquotedblright\ band.
For a sufficiently small interband coupling, one of the coherence lengths has
a peak near $T_{c,2}$. This peaked behavior of the coherence length and
related quantities was also considered in Ref. \cite{KomendovaPRL108}, where
it was referred to as \textquotedblleft hidden criticality\textquotedblright.
The peaked behavior near $T_{c,2}$ is most clearly revealed in the $\gamma$
susceptibility of the order parameters, $\partial\Psi_{j}/\partial\gamma$.
There is no true criticality in a two-band fermion system at $T<T_{c}$.
Rather, the terminology introduced in Ref. \cite{KomendovaPRL108} emphasizes
the fact that the coupled system is still affected by the proximity of the
weaker band.

\begin{figure}
[h]
\begin{center}
\includegraphics[
height=2.3947in,
width=3.0515in
]%
{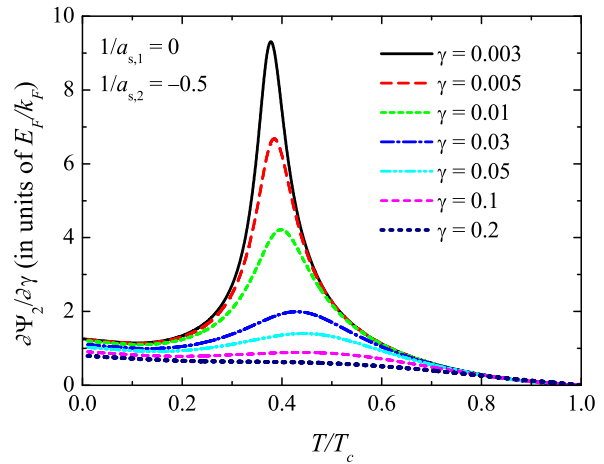}%
\caption{Susceptibility $\partial\Psi_{2}/\partial\gamma$ for the
\textquotedblleft weaker\textquotedblright\ band as a function of $T/T_{c}$
for inverse scattering lengths $1/\left(  k_{F}a_{s,1}\right)  =0$ and
$1/\left(  k_{F}a_{s,2}\right)  =-0.5$. Here, $T_{c}$ is the critical
temperature for the two-band system.}%
\label{fig:Gamma}%
\end{center}
\end{figure}

We find that the non-monotonic temperature dependence of the thermodynamic
quantities is also present in two-bandgap superfluid atomic Fermi gases, as
shown in Fig.~\ref{fig:Gamma}. For sufficiently weak interband coupling, a
peak appears in $\partial\Psi_{2}/\partial\gamma$ at $T\approx T_{c,2}$. Note
that in the standard GL model a non-physical divergence of $\partial\Psi
_{2}/\partial\gamma$ occurs at $T_{c,2}$: in order to find a finite
susceptibility peak the Bogoliubov - de Gennes (BdG) equations had to be used
in Ref. \cite{KomendovaPRL108}. We find that our present formalism, like the
BdG equations, leads to a convergent susceptibility at $T\approx T_{c,2}$.%

As already indicated above, a common chemical potential for the two bands can
lead to a partial depletion of the population of the \textquotedblleft
weak\textquotedblright\ band. This will also affect the critical temperature
$T_{c,2}$ corresponding to the weak band, even at zero interband coupling,
$\gamma=0$. This effect, to the best of our knowledge, did not attract
attention in past works on multiband superconductors, as these consider only
the BCS limit for the scattering lengths, where the feedback of the gap
parameter to the density and to the number equations is negligible.%

\begin{figure}
[h]
\begin{center}
\includegraphics[
height=2.4962in,
width=3.0607in
]%
{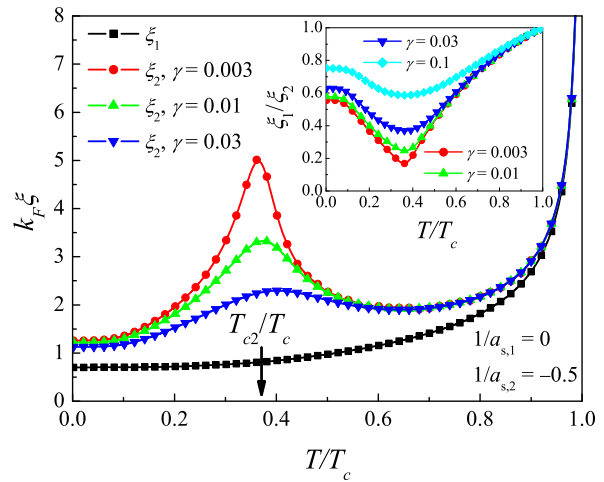}%
\caption{The healing lengths $\xi_{j}$ corresponding to the stronger ($j=1$)
and weaker ($j=2$) bands are shown as a function of temperature for different
coupling parameters $\gamma$.\ The inset shows the ratio $\xi_{1}/\xi_{2}$.}%
\label{fig:HL}%
\end{center}
\end{figure}

Having obtained the bulk values $\Psi_{1}$ and $\Psi_{2}$ for a uniform (bulk)
system, the healing lengths can be determined for the two-bandgap case by
substituting $\Psi_{j}(r)=\Psi_{j}^{(bulk)}\tanh(x/\sqrt{2}\xi_{j})$ in the
EFT energy functional for a stationary pair field. This variational
\textquotedblleft trial function\textquotedblright\ describes how a
two-bandgap superfluid in a semi-infinite space heals from a wall at $x=0$
back to the bulk values $\Psi_{j}^{(bulk)}$ of the band gaps. The healing
lengths $\xi_{j}$ are determined variationally and the result is shown in
Fig.~\ref{fig:HL}. We find, in agreement with Ref. \cite{KoganPRB83}, that the
ratio of healing lengths converges to 1 in the limit $T\rightarrow T_{c}$. The
obtained peaked behavior of the healing length $\xi_{2}$ near $T_{c,2}$ also
agrees with the results of Refs. \cite{Babaev2012-2,KomendovaPRL108} derived
using a very different method. These healing lengths will also determine the
structure of vortices in the fermionic superfluids. In superconductors, the
London penetration depth comes into play as a second length scale, but the
experiments on quantum gases work with neutral atoms so that there is no
coupling to the vector potential.

The healing length calculated here from the vortex profiles should be
distinguished from the pair correlation length as discussed in Ref.
\cite{Palestini}. The latter should be calculated from the correlation
function $g_{\uparrow\downarrow}\left(  \rho\right)  $ as in Ref.
\cite{Palestini}.

\subsection{Spectra of collective modes \label{subsec3c}}

In a two-band system, the Bogoliubov -- Anderson (Goldstone) collective mode
should also exist, as in a one-band system. The existence of a Goldstone mode
is a universal result, caused by the spontaneous breakdown of gauge symmetry
associated with the superfluid phase transition. In addition, another
collective mode can appear in a two-band system, as first derived by A.
Leggett \cite{Leggett} for a two-band BCS superconductor. The Leggett mode has
been observed in MgB$_{2}$ using Raman scattering \cite{Raman}, but to the
best of our knowledge, it has not yet been observed in multi-component atomic
Fermi gases. This has not stopped theoretical efforts to consider two-band
Fermi superfluids in the BCS-BEC crossover regime at zero temperature
\cite{Iskin,Iskin2}. In Refs. \cite{Iskin,Iskin2} the model interband
Josephson interaction is present already in the starting Hamiltonian, and the
interaction is measured in terms of potentials rather than scattering lengths.
In the present treatment, the Josephson interaction emerges from the
interatomic scattering interactions. Nevertheless, we can perform a
qualitative comparison of our results with those of Refs. \cite{Iskin,Iskin2}.

The spectrum of collective excitations in a one-band system was determined
solving the equation (\ref{det}) where the dynamic matrix $\mathbb{M}\left(
q,\omega\right)  $ is given by (\ref{M}). As follows from the effective action
(\ref{Squad}), the dynamic matrix for the two-band system is a $4\times4$
matrix that can be written as%
\begin{equation}
\mathbb{M}^{\left(  2band\right)  }\left(  q,\omega\right)  =\left(
\begin{array}
[c]{cc}%
\mathbb{M}_{1}\left(  q,\omega\right)  +\varkappa\eta\cdot\mathbb{I} &
-\varkappa\mathbb{I}\\
-\varkappa\mathbb{I} & \mathbb{M}_{2}\left(  q,\omega\right)  +\frac
{\varkappa}{\eta}\cdot\mathbb{I}%
\end{array}
\right)  , \label{M2b}%
\end{equation}
where $\mathbb{M}_{j}\left(  q,\omega\right)  $ are the dynamic matrices for
each band, $\mathbb{I}$ is the unit $2\times2$ matrix, $\varkappa$ is
proportional to the interband coupling strength $\varkappa\equiv\frac
{m_{12}\gamma}{4\pi}$, and $\eta=\frac{\left\vert \Psi_{2}\right\vert
}{\left\vert \Psi_{1}\right\vert }$ is the ratio of the gap parameters for two
bands. The eigenfrequencies of the collective excitations are determined by
the roots of the equation%
\begin{equation}
\det\mathbb{M}^{\left(  2band\right)  }\left(  q,\omega\right)  =0.
\label{detM}%
\end{equation}

We are searching for the eigenfrequencies at small momenta $q$, that is in
accordance with the approximation of slowly varying pair fields. Therefore the
roots of the equation (\ref{detM}) are approximated by the leading terms of
the Taylor series in powers of the momentum, similarly as for the one band system.

\paragraph{Bogoliubov -- Anderson mode}

The Bogoliubov -- Anderson mode at small $q$ is an acoustic mode $\omega
_{q}=v_{s}q$ determined by the sound velocity $v_{s}$. For the two-band system
we find:%
\begin{align}
v_{s}  &  =\left\{  \left(  \varkappa\left(  \mathcal{U}_{1}+\eta
^{2}\mathcal{U}_{2}\right)  +2\eta\mathcal{U}_{1}\mathcal{U}_{2}\right)
\left(  \frac{\mathcal{\tilde{C}}_{1}}{m_{1}}+\eta^{2}\frac{\mathcal{\tilde
{C}}_{2}}{m_{2}}\right)  \right. \nonumber\\
&  \times\left[  \varkappa\left(  \mathcal{\tilde{D}}_{1}+\eta^{2}%
\mathcal{\tilde{D}}_{2}\right)  ^{2}+2\eta\left(  \mathcal{\tilde{D}}_{1}%
^{2}\mathcal{U}_{2}+\eta^{2}\mathcal{\tilde{D}}_{2}^{2}\mathcal{U}_{1}\right)
\right. \nonumber\\
&  \left.  \left.  +2\left(  \varkappa\mathcal{U}_{1}+\varkappa\eta
^{2}\mathcal{U}_{2}+2\eta\mathcal{U}_{1}\mathcal{U}_{2}\right)  \left(
\mathcal{\tilde{Q}}_{1}+\eta^{2}\mathcal{\tilde{Q}}_{2}\right)  \right]
^{-1}\right\}  ^{1/2}. \label{vs2b}%
\end{align}

\begin{figure}
[h]
\begin{center}
\includegraphics[
height=2.3514in,
width=3.0322in
]%
{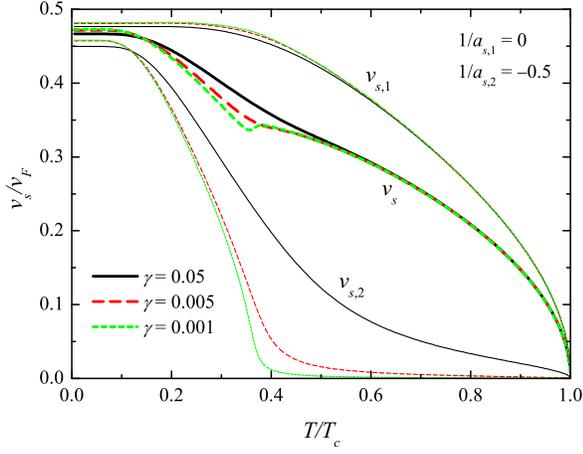}%
\caption{\emph{Heavy curves}: The sound velocity in a two-band superfluid
Fermi gas as a function of the temperature for $1/a_{s,1}=0$, $1/a_{s,2}%
=-0.5,$ with different values of the interband coupling parameter $\gamma$.
\emph{Thin curves}: the one-band sound velocity parameters $v_{s,1}$ and
$v_{s,2}$ described in the text.}%
\label{fig:vs2}%
\end{center}
\end{figure}

In Fig.~\ref{fig:vs2}, the temperature dependence of the sound velocity in a
two-band system is shown. We plot the sound velocity as a function of
temperature for $1/a_{s,1}=0$ and $1/a_{s,2}=-0.5,$ using different values of
the interband coupling parameter $\gamma$. For comparison, the one-band sound
velocities for each band are shown in the same graph by thin curves. They are
calculated using formula (\ref{vs}) with the parameters $\left(  \beta
,\mu,\left\vert \Psi\right\vert \right)  $ attributed to each band in the
\emph{coupled} two-band system (rather than with the parameters for an
independent one-band system). In other words, in the figure, $v_{s,j}%
=v_{s}^{\left(  1band\right)  }\left(  \beta,\mu,\left\vert \Psi
_{j}\right\vert \right)  $, where $\left\vert \Psi_{j}\right\vert $ are
determined from the coupled gap equations (\ref{LDA2b}) for the two-band
system with the number equation (\ref{NumEq}). Under these conditions, the
inequality%
\[
\min\left(  v_{s,1},v_{s,2}\right)  \leq v_{s}\leq\max\left(  v_{s,1}%
,v_{s,2}\right)
\]
is fulfilled. It should be noted that $v_{s,1},v_{s,2}$ are not the true sound
velocities in the two-band system: they are only auxiliary parameters. There
is a unique first sound velocity $v_{s}$ for the whole system given by
(\ref{vs2b}).%

We see from Fig.~\ref{fig:vs2} that $v_{s,2}$ shows a rapid decrease for
temperatures near $T_{c,2}<T_{c}$. This decrease of $v_{s,2}$ becomes more
gradual as the interband coupling $\gamma$ becomes larger. This is reflected
in the behavior of $v_{s}(T)$: near $T_{c}$ the sound velocity of the
Bogoliubov -- Anderson mode shows a dip.

\paragraph{Leggett mode}

The Leggett mode is specific for two-band superfluids as it describes small
oscillations of the relative phase of two condensates. In the long-wavelength
approximation, the frequency of the Leggett mode can be approximately written
as%
\begin{equation}
\omega_{L}\left(  q\right)  \approx\sqrt{\omega_{L,0}^{2}+v_{L}^{2}q^{2}},
\label{wL}%
\end{equation}
so that it remains gapped in the limit $q\rightarrow0$ (for $\gamma\neq0$).
The frequency of the Leggett mode is determined numerically by solving Eq.
(\ref{detM}). In the long-wavelength approximation an analytic approximation
for the Leggett mode frequency can be obtained:%
\begin{equation}
\omega_{L,0}\approx\left(  \frac{2\mathcal{U}_{1}\varkappa\eta}{P_{1}%
^{2}+2\mathcal{U}_{1}\mathcal{\tilde{Q}}_{1}}+\frac{2\mathcal{U}_{1}%
\frac{\varkappa}{\eta}}{P_{2}^{2}+2\mathcal{U}_{2}\mathcal{\tilde{Q}}_{2}%
}\right)  ^{1/2}. \label{wL0}%
\end{equation}

\begin{figure}
[h]
\begin{center}
\includegraphics[
height=2.2748in,
width=3.0137in
]%
{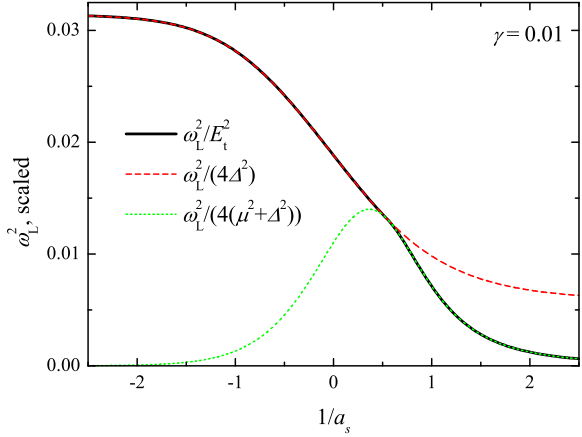}%
\caption{The frequency (squared) of the Leggett mode for a two-band Fermi gas
at $T=0$ is shown as a function of $1/a_{s}$ for the interband coupling
$\gamma=0.01$. The Leggett mode frequency is scaled by the two-quasiparticle
threshold $E_{t}\equiv\min\left(  2E_{\mathbf{k}}\right)  $ (full line),
$2\Delta$ (thin dashed line) and $2\sqrt{\Delta^{2}+\mu^{2}}$ (thin dot-dashed
line), respectively.}%
\label{fig:vs3}%
\end{center}
\end{figure}

In Fig.~\ref{fig:vs3}, we plot the frequency (squared) of the Leggett mode,
$\omega_{L,0}^{2}$, for a two-band Fermi gas at $T=0$, as a function of the
inverse scattering length $1/a_{s}=1/a_{s,1}=1/a_{s,2}$. The Leggett mode
frequency is scaled to the two-particle threshold $E_{t}\equiv\min\left(
2E_{\mathbf{k}}\right)  $ similarly as in Ref. \cite{Iskin} where $E_{t}$ is
deemed to be the physically reasonable maximal value for the frequency of
collective oscillations. This scaling factor is equal to $2\Delta$ for $\mu>0$
and $2\sqrt{\Delta^{2}+\mu^{2}}$ for $\mu<0$, where $\Delta=\max\left(
\left\vert \Psi_{1}\right\vert ,\left\vert \Psi_{2}\right\vert \right)  $.%

We can qualitatively compare the behavior of $\omega_{L,0}^{2}$ obtained in
the present treatment with the result shown in Fig.~5 of Ref. \cite{Iskin}. At
present, it is not obvious how the parameters used in the theory of Refs.
\cite{Iskin,Iskin2} can be matched to the scattering lengths used in our
approach. There is a difference between the two theories, because the starting
atomic Hamiltonian of Ref. \cite{Iskin2} contains \emph{a priori} scattering
between Cooper pairs, which is not invoked in the present formalism.
Nevertheless, we can see a clear similarity between the behavior of
$\omega_{L,0}^{2}$ in Ref. \cite{Iskin} and in the present treatment.%

\begin{figure}
[h]
\begin{center}
\includegraphics[
height=4.1382in,
width=2.9648in
]%
{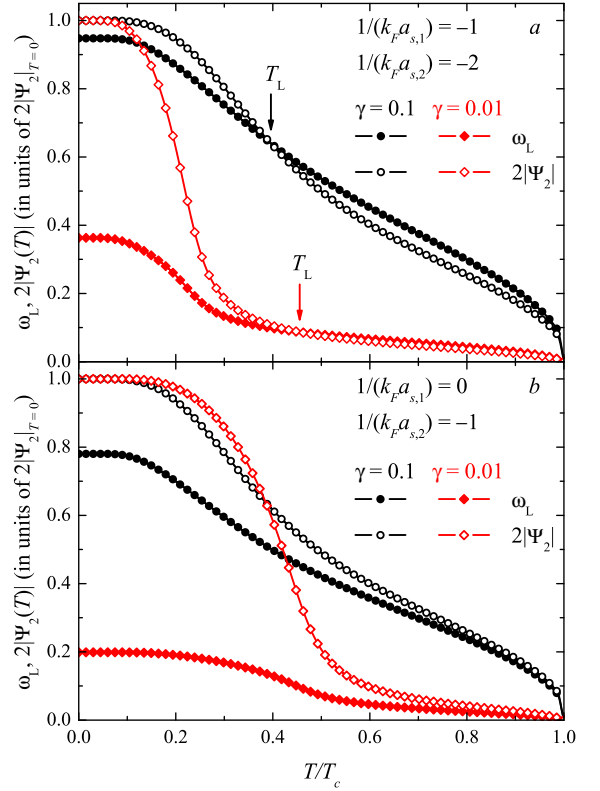}%
\caption{Temperature dependence of the Leggett mode frequencies $\omega_{L}$
(full symbols) and twice the order parameter $\left\vert \Psi_{2}\right\vert $
(hollow symbols) for $1/a_{s,1}=-1$, $1/a_{s,2}=-2$ (\emph{a}) and for
$1/a_{s,1}=0$, $1/a_{s,2}=-1$ (\emph{b}), at different values of the interband
coupling strength, $\gamma=0.1$ (circles) and $\gamma=0.01$ (diamonds).}%
\label{fig:vs4}%
\end{center}
\end{figure}

In Fig.~\ref{fig:vs4}, the temperature dependence of the Leggett mode
frequency is analyzed for different values of the intraband scattering lengths
and of the interband coupling strength. The Leggett mode softens with
increasing temperature and turns to zero at $T=T_{c}$. The Leggett mode cannot
exist in a one-band system, because it describes oscillations of the relative
phase of two condensates. Therefore the Leggett mode frequency must drop down
as $T>T_{c,2}$, especially at weak interband coupling. This trend is clearly
visible in Fig.~\ref{fig:vs4}. The behavior of the Leggett mode as a function
of temperature obtained in the present formalism agrees qualitatively well
with the experimental measurement \cite{Ponomarev} of this mode in MgB$_{2}$,
and with different theoretical approaches \cite{Maksimov,Ichioka}.

The observation of the Leggett mode in two-band superconductors was
problematic during a long time, because at $\omega_{L}>2\left\vert \Psi
_{2}\right\vert $, the Leggett mode can dissipate to one-particle excitations
\cite{Leggett}. Therefore the range of observation for the Leggett mode is
approximately restricted by the condition $\omega_{L}<2\left\vert \Psi
_{2}\right\vert $. As pointed out in \cite{Sharapov}, the weak-band order
parameter in conventional superconductors is very small, thus making the
experimental observation of the Leggett mode rather difficult. However,
recently the Leggett mode has been clearly indicated
\cite{Raman,Ponomarev,Maksimov}. In Fig.~\ref{fig:vs4}, the temperature
dependence of the Leggett mode frequencies is shown for two sets of values of
the intraband scattering lengths: the BCS regime for both bands with
$1/a_{s,1}=-1,\ 1/a_{s,2}=-2$ (panel \emph{a}) and the case with
$1/a_{s,1}=0,\ 1/a_{s,2}=-1$ (panel \emph{b}). In order to see the range for
the possible experimental observation of the Leggett mode, twice the order
parameter for a \textquotedblleft weak\textquotedblright\ band is plotted in
the same figure. The arrows indicate the upper bound temperature $T_{L}$ for
the observation of the Leggett mode, where $\omega_{L}=2\left\vert \Psi
_{2}\right\vert $. At $T>T_{L}$, the Leggett mode dissipates, and at $T<T_{L}$
it can be observable. As seen from Fig.~\ref{fig:vs4} (\emph{a}), this upper
bound temperature exists in the BCS regime. For a stronger coupling
$1/a_{s,1}=0,\ 1/a_{s,2}=-1$, however, the condition $\omega_{L}<2\left\vert
\Psi_{2}\right\vert $ is fulfilled in the whole range $0<T<T_{c}$. We can
conclude that the strong-coupling regime is more favorable for the observation
of the Leggett mode than the BCS regime. The strong-coupling regime has been
experimentally realized in the condensed atomic gases using the Feshbach
resonance. Therefore the observation of the Leggett mode in ultracold Fermi
gases is expectable.

\section{Conclusions \label{Conclusions}}

The first main result of the present work is the derivation of a finite
temperature, all-coupling effective field theory for superfluid Fermi gases,
obtained by performing a gradient expansion of the pair field around a
background value that is not necessarily small. Assuming the validity of the
derivative expansion for the order parameter, the effective field action
functional has been obtained by systematically summing \emph{all} terms in
powers of the order parameter, and is therefore valid at all temperatures
below $T_{c}$. The expansion has been performed up to second order in both
spatial gradients and time derivatives, so that the resulting effective field
theory is capable of describing collective excitations for temperatures below
$T_{c}$. The finite-temperature EFT is a straightforward extension of several
preceding approaches: the effective field theory developed for $T\approx
T_{c}$ \cite{SadeMeloPRL71,HuangPRA79} and that developed for $T=0$
\cite{Marini1998,Schakel}. The current formalism corresponds with these
approaches in the appropriate limiting cases. Also we retrieve the BCS-BEC
theory result for the ground state energy at $T=0$. Finally, the results for
vortices (described here) and for solitons (described in Ref \cite{DSol})
correspond well with the results obtained from Bogoliubov - de Gennes
calculations. The advantage of the current formalism is that the coefficients
of the proposed action functional (\ref{SEFT}) are closed and tractable
expressions, which turn to the known GL coefficients in the limit
$T\rightarrow T_{c}$, and are fast to compute.

The present EFT describes the ultracold Fermi gases in the BCS-BEC crossover,
smoothly passing the unitarity regime, similarly to Ref. \cite{SadeMeloPRL71}
and the related analytic theories. The unitarity regime needs however a
special care, as discussed in Sect. \ref{EFT}. Accurate quantitative results
have been obtained at unitarity and $T=T_{c}$ using numeric approaches
\cite{Bulgac,Burovski}, and analytically, using different methods, e. g.
$\varepsilon$-expansion \cite{Nishida}, $1/N$-expansion \cite{Enss}, the
renormalization group methods \cite{Nikolic,Gubbels,Boettcher}. However, these
analytic methods are focused at the unitatity regime. At present, to the best
of our knowledge, there is no known analytic theory which accurately
quantitatively describes the Fermi gases in a unified way through the whole
BCS-BEC crossover, including the unitarity point. Therefore the analytic
theories smoothly describing the ultracold Fermi gases in the BCS-BEC
crossover are useful, because they can provide a reasonable description of
ultracold Fermi gases in the whole range of the coupling strength, except the
aforesaid singularity point.

It is established in Ref. \cite{SadeMeloPRL71} that the solution for the
critical temperature obtained within the functional integral method accounting
for Gaussian fluctuations about the saddle point smoothly interpolates between
the two limiting cases -- BCS and BEC regimes. Many subsequent works use
analytic approximations similar to that in Ref. \cite{SadeMeloPRL71} for the
thermodynamic functions of ultracold Fermi gases (e. g.
\cite{HuangPRA79,Diener2008,Haussmann}). Also our work follows this direction,
being particularly aimed on the treatment of the ultracold Fermi gases below
$T_{c}$.

\bigskip

The second main result of this work is the extension of the effective field
formalism to the case of two-band fermionic superfluids. The resulting
effective field action contains the same input parameters (the scattering
lengths and the interband coupling strength) as those in other approaches to
the two-band superfluidity/superconductivity, e. g. Refs.
\cite{SuhlPRL3,Iskin,Iskin2,Iskin3,ZhitomirskyPRB69}. These input parameters
can be independently measured and even precisely controlled -- for ultracold
gases. They completely fix the microscopic Hamiltonian for the mixture of two
atomic Fermi superfluids, with Cooper pairing within each superfluid and
contact interactions between the atoms belonging to different superfluids. In
the effective bosonic action obtained by the path-integral treatment of this
Hamiltonian, the two superfluid order parameters are coupled by a Josephson
term that is not introduced \emph{ad hoc}, but follows directly from applying
the Hubbard--Stratonovich transformation, before any approximation is made.
This Josephson coupling is kept also after performing the gradient expansion
which results in the effective two-field theory.

For the two-band superfluid, the current theory reveals a non-monotonic
temperature behavior of the thermodynamic parameters near the (uncoupled)
critical temperature of a \textquotedblleft weaker\textquotedblright\ band,
similar to that obtained for two-band superconductors with a Bogoliubov - de
Gennes treatment (whereas the standard Ginzburg - Landau treatment fails to
reproduce this). Also the existence of two healing length scales is captured
by the present effective field theory for a two-band Fermi superfluid.

The formalism developed here can find a broad spectrum of applications, in
particular as a complementary method to the Bogoliubov - de Gennes equations
which are restricted to the mean-field approximation and to the BCS case, and
which become cumbersome when many vortices are present. The present method is
applicable in the whole range of the BCS-BEC crossover, allows one to take
into account the fluctuations, and requires much less computation time than
the Bogoliubov - de Gennes formalism. Moreover, the EFT allows one to obtain
analytical solutions in some cases where the Bogoliubov - de Gennes equations
can be solved only numerically.

\begin{acknowledgments}
The authors gratefully acknowledge support of the Research Fund of the
University of Antwerp. This work was supported by FWO-V projects G.0429.15N,
G.0370.09N, G.0180.09N, G.0115.12N, G.0119.12N, the WOG WO.033.09N (Belgium).
\end{acknowledgments}

\appendix

\section{Effective bosonic action for a two-band system}

The integration over the fermionic variables in (\ref{Z1}) is performed
exactly. We use the Nambu representation with four-dimensional spinors%
\begin{equation}
\psi=\left(
\begin{array}
[c]{c}%
\psi_{1,\uparrow}\\
\bar{\psi}_{1,\downarrow}\\
\psi_{2,\downarrow}\\
\bar{\psi}_{2,\uparrow}%
\end{array}
\right)  . \label{Nambu2}%
\end{equation}
Note that for the second band we use spin projections opposite to those used
in the first band. The HS action (\ref{SHS}) is then represented in matrix
form as follows:%
\begin{align}
S_{HS}  &  =S_{B}+S_{\chi}\nonumber\\
&  +\frac{1}{2}\int_{0}^{\beta}d\tau\int d\mathbf{r}\left(
\begin{array}
[c]{cc}%
\psi & \bar{\psi}%
\end{array}
\right) \nonumber\\
&  \times\left(
\begin{array}
[c]{cc}%
\left(  -\mathbb{G}^{-1}\right)  _{1,1} & \left(  -\mathbb{G}^{-1}\right)
_{1,2}\\
\left(  -\mathbb{G}^{-1}\right)  _{2,1} & \left(  -\mathbb{G}^{-1}\right)
_{2,2}%
\end{array}
\right)  \left(
\begin{array}
[c]{c}%
\psi\\
\bar{\psi}%
\end{array}
\right)  , \label{SHS1}%
\end{align}
with $S_{B}$ and $S_{\chi}$ given by (\ref{SB}) and (\ref{Sxi}), respectively.
The following matrices for the inverse 4-dimensional Nambu tensor are
introduced:%
\begin{equation}
\left(  -\mathbb{G}^{-1}\right)  _{1,1}=\left(
\begin{array}
[c]{cccc}%
0 & 0 & 0 & -i\chi_{2}\\
0 & 0 & i\chi_{2} & 0\\
0 & -i\chi_{2} & 0 & 0\\
i\chi_{2} & 0 & 0 & 0
\end{array}
\right)  , \label{aa11}%
\end{equation}%
\begin{align}
&  \left(  -\mathbb{G}^{-1}\right)  _{1,2}\nonumber\\
&  =\left(
\begin{array}
[c]{cccc}%
\frac{\partial}{\partial\tau}-\hat{H}_{1} & \bar{\Psi}_{1} & -i\chi_{1} & 0\\
\Psi_{1} & \frac{\partial}{\partial\tau}+\hat{H}_{1} & 0 & i\chi_{1}\\
-i\bar{\chi}_{1} & 0 & \frac{\partial}{\partial\tau}-\hat{H}_{2} & -\bar{\Psi
}_{2}\\
0 & i\bar{\chi}_{1} & -\Psi_{2} & \frac{\partial}{\partial\tau}+\hat{H}_{2}%
\end{array}
\right)  , \label{aa12}%
\end{align}%
\begin{align}
&  \left(  -\mathbb{G}^{-1}\right)  _{2,1}\nonumber\\
&  =\left(
\begin{array}
[c]{cccc}%
\frac{\partial}{\partial\tau}+\hat{H}_{1} & -\Psi_{1} & i\bar{\chi}_{1} & 0\\
-\bar{\Psi}_{1} & \frac{\partial}{\partial\tau}-\hat{H}_{1} & 0 & -i\bar{\chi
}_{1}\\
i\chi_{1} & 0 & \frac{\partial}{\partial\tau}+\hat{H}_{2} & \Psi_{2}\\
0 & -i\chi_{1} & \bar{\Psi}_{2} & \frac{\partial}{\partial\tau}-\hat{H}_{2}%
\end{array}
\right)  , \label{aa21}%
\end{align}%
\begin{equation}
\left(  -\mathbb{G}^{-1}\right)  _{2,2}=\left(
\begin{array}
[c]{cccc}%
0 & 0 & 0 & i\bar{\chi}_{2}\\
0 & 0 & -i\bar{\chi}_{2} & 0\\
0 & i\bar{\chi}_{2} & 0 & 0\\
-i\bar{\chi}_{2} & 0 & 0 & 0
\end{array}
\right)  . \label{aa22}%
\end{equation}
The integration over the fermion fields $\psi\,\ $is performed in the same way
as in Ref. \cite{SadeMeloPRL71} and results in a partition function written as
a the path integral over the boson fields $\Psi$ and $\chi$,%
\begin{equation}
\mathcal{Z}\propto\int\mathcal{D}\left[  \bar{\Psi},\Psi\right]
\int\mathcal{D}\left[  \bar{\chi},\chi\right]  e^{-S_{eff}}, \label{Z2}%
\end{equation}
where the effective bosonic action depends on the pair and density fields
through%
\begin{equation}
S_{eff}=S_{B}-\sum_{j=1,2}\operatorname{Tr}\ln\left[  -\mathbb{G}_{j}%
^{-1}\right]  . \label{Seff1}%
\end{equation}

\end{document}